\input form
\input chrform

\def\chr{{\it X$\rho$\'o$\nu o\varsigma$}}
\overfullrule=0pt
\cita{Birkhoff-1927}{Birkhoff, G. D.: {\it Dynamical systems}, New York 
(1927).}

\cita{Biscani-2009}{Biscani, F.: {\it The Piranha algebraic
manipulator}, arXiv:0907.2076 (2009).}

\cita{Broucke-1969}{Broucke, R. and Garthwaite, K.: {\it A Programming System 
for Analytical Series Expansions on a Computer} Cel. Mech., {\bf 1},
271--284 (1969).}

\cita{Broucke-1989}{Broucke, R.: {\it A Fortran-based Poisson series processor 
and its applications in celestial mechanics}, Cel. Mec., {\bf 45},
255--265 (1989).}

\cita{Contopoulos-1960}{Contopoulos, G.: {\it A third integral of motion in a 
Galaxy}, Z. Astrophys., {\bf 49}, 273--291 (1960).}

\cita{Contopoulos-2002}{Contopoulos, G.: {\it Order and Chaos in
Dynamical Astronomy}, Springer (2002).}

\cita{Contopolus-2003}{Contopoulos, G., Efthymiopoulos, C. and Giorgilli, A.:
{\it Non-convergence of formal integrals of motion}, J. Phys. A:
Math. Gen., {\bf 36}, 8639--8660 (2003).}

\cita{Contopolus-2004}{Contopoulos, G., Efthymiopoulos, C. and Giorgilli, A.:
{\it Non-convergence of formal integrals of motion II: Improved
Estimates for the Optimal Order of Truncation}, J. Phys. A:
Math. Gen., {\bf 37}, 10831--10858 (2004).}

\cita{Giorgilli-1978}{Giorgilli, A., Galgani, L.: {\it Formal
integrals for an autonomous Hamiltonian system near an equilibrium
point}, Cel. Mech., {\bf 17}, 267--280 (1978).}

\cita{Giorgilli-1979}{Giorgilli, A.: {\it A computer program for integrals
of motion}, Comp. Phys. Comm., {\bf 16}, 331--343 (1979).}

\cita{Giorgilli-1988.4}{Giorgilli, A.: {\it Rigorous results on the 
power expansions for the integrals of a Hamiltonian system near an
elliptic equilibrium point}, Ann. Ist. H. Poincar\'e, {\bf 48},
423--439 (1988).}

\cita{Giorgilli-1989}{Giorgilli, A., Delshams, A., Fontich, E., Galgani, L. and 
Sim\'o, C.: {\it Effective stability for a Hamiltonian system near an 
elliptic equilibrium point, with an application to the restricted three 
body problem.} J. Diff. Eqs., {\bf 20}, (1989).
}

\cita{Giorgilli-1997}{Giorgilli, A. and Skokos, Ch.:{\it On the
stability of the Trojan asteroids}, Astron. Astroph., {\bf 317},
254--261 (1997).}

\cita{Gustavson-1966}{Gustavson, F. G.: {\it On constructing formal
integrals of a Hamiltonian system near an equilibrium point}, Astron.
J., {\bf 71}, 670--686 (1966).}

\cita{Henon-1964}{H\'enon, M. and Heiles, C.: {\it The applicability of 
the third integral of motion: some numerical experiments}, Astron. J., 
{\bf 69}, 73--79 (1964).}

\cita{Henrard-1974}{Henrard, J.: {\it Equivalence for Lie transforms}, 
Cel. Mech., {\bf 10}, 497--512 (1974).}

\cita{Henrard-1986}{Henrard, J.: {\it Algebraic manipulations on 
computers for lunar and planetary theories}, Relativity in celestial
mechanics and astrometry, 59--62 (1986)}

\cita{Henrard-1989}{Henrard, J.: {\it A survey on Poisson series
processors}, Cel. Mech., {\bf 45}, 245--253 (1989).}

\cita{Jorba-1999}{Jorba, J.: {\it A Methodology for the Numerical
Computation of Normal Forms, Centre Manifolds and First Integrals of
Hamiltonian Systems}, Experimental Mathematics, {\bf 8}, 155--195 (1999).}

\cita{Knuth-1968}{Knuth, D.: {\it The Art of Computer Programming}, 
Addison-Wesley (1968)}

\cita{Laskar-1989.1}{Laskar, J: {\it Manipulation des s\'eries}, in
D.\ Benest and C.\ Froeschl\'e (eds.), {\it Les m\'ethodes modernes
de ma m\'echanique c\'eleste (Goutelas, 1989)}, Ed.\ Fonti\`eres,
89--107 (1989).}

\cita{Poincare-1889}{Poincar\'e, H.: {\it Sur le probl\`eme des trois
corps et les \'equations de la dynamique},  Acta Mathematica (1890).}

\cita{Poincare-1892}{Poincar\'e, H.: {\it Les m\'ethodes nouvelles de la 
m\'ecanique c\'eleste}, Gauthier--Villars, Paris (1892).
}

\cita{Rom-1970}{Rom, A.: {\it Mechanized Algebraic Operations (MAO)}, 
Cel. Mech., {\bf 1}, 301--319 (1970).}

\cita{Sansottera-2010}{Sansottera, M., Locatelli U., and Giorgilli, A.: 
{\it On the stability of the secular evolution of the
planar Sun-Jupiter-Saturn-Uranus system}, Math. Comput. Simul.,
doi:10.1016/j.matcom.2010.11.018.}

\cita{Siegel-1941}{Siegel, C. L.: {\it On the integrals of canonical 
systems}, Ann. Math., {\bf 42}, 806--822 (1941).}

\cita{Whittaker-1916}{Whittaker, E. T.: {\it On the adelphic integral of the 
differential equations of dynamics}, Proc. Roy Soc. Edinburgh, Sect. A, 
{\bf 37}, 95--109 (1916).}


\title{Methods of algebraic manipulation \hfil\break in perturbation theory}

\author{\it ANTONIO GIORGILLI 
\hfill\break Dipartimento di Matematica, Universit\`a degli Studi di Milano,
\hfill\break Via Saldini 50, 20133\ ---\ Milano, Italy
\hfill\break and
\hfill\break Istituto Lombardo Accademia di Scienze e Lettere}

\author{\it MARCO SANSOTTERA
\hfill\break naXys, Namur Center for Complex Systems, FUNDP,
\hfill\break Rempart de la Vierge 8,  B5000 \ ---\ Namur, Belgium.}

\abstract{We give a short introduction to the methods of representing
polynomial and trigonometric series that are often used in Celestial
Mechanics.  A few applications are also illustrated.}

\section{1}{Overview}
Algebraic manipulation on computer is a tool that has been developed
quite soon, about one decade after the birth of computers, the first
examples dating back to the end of the fifties of the last century.
General purpose packages began to be developed during the sixties, and
include, e.g., Reduce (1968), Macsyma (1978), muMath (1980), Maple
(1984), Scratchpad (1984), Derive (1988), Mathematica (1988), Pari/GP
(1990) and Singular (1997) (the dates refer to the first release).
However, most of the facilities of these general purpose manipulators
are simply ignored when dealing with perturbation methods in Celestial
Mechanics.  For this reason, the job of developing specially devised
manipulation tools has been undertaken by many people, resulting in
packages that have limited capabilities, but are definitely more
effective in practical applications.  Producing a list of these
packages is a hard task, mainly because most of them are not publicly
available.  A list of ``old time'' packages may be found in
Henrard~\dbiref{Henrard-1989} and Laskar~\dbiref{Laskar-1989.1}.  In
recent times a manipulator developed by J.\ Laskar and M.\ Gastineau
has become quite known.

Finding references to the methods implemented in specially devised
packages is as difficult as giving a list.  We know only a few papers
by Broucke and Garthwaite~\dbiref{Broucke-1969},
Broucke~\dbiref{Broucke-1989}, Rom~\dbiref{Rom-1970},
Henrard~\dbiref{Henrard-1986} and~\dbiref{Henrard-1989},
Laskar~\dbiref{Laskar-1989.1}, Jorba~\dbiref{Jorba-1999} and
Biscani~\dbiref{Biscani-2009}.  A complete account of the existing
literature on the subject goes beyond the limits of the present note.
The present work introduces some ideas that have been used by the
authors in order to implement a package named \chr.

As a matter of fact, most of the algebraic manipulation used in
Celestial Mechanics makes use of the so called ``Poisson series'',
namely series with a general term of the form
$$
x_1^{j_1}\cdot\ldots\cdot x_n^{j_n} 
{{\cos}\atop{\sin}} (k_1\phi_1+\ldots+k_m\phi_m)\ ,
$$
(with obvious meaning of the symbols).  Thus, a very minimal set of
operations is required, namely sums, products and derivatives of
polynomials and/or trigonometric polynomials.  Traditionally, also the
operation of inversion of functions, usually made again via series
expansion, was required.  However, the expansion methods based on Lie
series and Lie transforms typically get rid of the latter operation
(see, e.g.,~\dbiref{Henrard-1974}).

Writing a program doing algebraic manipulation on series of the type
above leads one to be confronted with a main question, namely how to
represent a polynomial, trigonometric polynomial or Poisson series on
a computer.  The papers quoted above actually deal with this problem,
suggesting some methods.  In these lectures we provide an approach to
this problem, followed by a few examples of applications.

In sect.~\secref{1.a} we include a brief discussion about
the construction of normal form for a Hamiltonian
system in the neighborhood of an elliptic equilibrium.  We do not
attempt to give a complete discussion, since it is available in many
papers.  We rather try to orient the reader's attention on the problem
of representing perturbation series.

In sect.~\secref{2}--\secref{4} we introduce a method which turns out
to be quite useful for the representation of a function as an array of
coefficients.  The basic idea has been suggested to one of the authors
by the paper of Gustavson~\dbiref{Gustavson-1966} (who, however, just
mentions that he used an indexing method, without giving any detail about
its implementation).  One introduces an {\corsivo indexing function}
which transforms an array of exponents in a polynomial (or
trigonometric polynomial) in a single index within an array.  The
general scheme is described in sect.~\secref{2}.  The basics behind
the construction of an indexing function are described in
sect.~\secref{3}.  The details concerning the representation of
polynomials and trigonometric polynomials are reported in
sects.~\secref{indexf.1} and~\secref{indexf.2}, respectively.  In
sect.~\secref{4} we include some hints about the case of sparse
series, that may be handled by combining the indexing functions above
with a tree representation.  Finally, sect.~\secref{5} is devoted to
three applications, by giving a short account of the contents of
published papers.

\section{1.a}{A common problem in perturbation theory}
A typical application of computer algebra is concerned with the
construction of first integrals or of a normal form for a Hamiltonian
system.  A nontrivial example, which however may be considered as a
good starting point, is the calculation of a normal form for the
celebrated model of H\'enon and Heiles~\dbiref{Henon-1964}, which has
been done by Gustavson~\dbiref{Gustavson-1966}.  Some results on this
model are reported in sect.~\secref{5}.  

We assume that the reader is not completely unfamiliar with the
concept of normal form for a (possibly Hamiltonian) system of
differential equations.  Thus, let us briefly illustrate the problem
by concentrating our attention on the algorithmic aspect and by
explaining how algebraic manipulation may be introduced.

\subsection{1.a.1}{Computation of a normal form} 
Let us consider a canonical system of differential equations in the
neighborhood of an elliptic equilibrium.  The Hamiltonian may
typically be given the form of a power series expansion
$$
H(x,y) = H_0(x,y) + H_1(x,y) +\ldots\ ,\quad
 H_0(x,y) = \sum_{j=1}^{n} \frac{\omega_j}{2}(x_j^2+y_j^2)\ ,
\formula{nrm.1} 
$$
where $H_{s}(x,y)$ for $s\ge 1$ is a homogeneous polynomial of degree
$s+2$ in the canonical variables $(x,y)\in\reali^{2n}$.  Here
$\omega\in\reali^n$ is the vector of the frequencies, that are assumed
to be all different from zero.

In such a case the system is said to be in Birkhoff normal form in
case the Hamiltonian takes the form
$$
H(x,y) = H_0(x,y) + Z_1(x,y) + Z_2(x,y)+\ldots\quad
{\rm with}\quad L_{H_0} Z_s =0\ ,
\formula{nrm.2}
$$
where $L_{H_0}\cdot = \{H_0,\cdot\}$ is the Lie derivative with respect
to the flow of $H_0$, actually the Poisson bracket with $H_0$.

The concept of Birkhoff normal form is better understood if one
assumes also that the frequencies are non resonant, i.e., if 
$$
\langle k,\omega\rangle \ne 0\quad {\rm for\ all\ } k\in\interi^n
\>,\ k\ne 0\ ,
$$
where $\langle k,\omega\rangle = \sum_{j}k_j\omega_j$. 
For, in this case the functions $Z_s(x,y)$ turn out to be actually
function only of the $n$ actions of the system, namely of the
quantities
$$
I_j = \frac{x_j^2+y_j^2}{2}\ ,\quad j=1,\ldots,n\ .
$$
It is immediate to remark that $I_1,\ldots,I_n$ are independent first
integrals for the Hamiltonian, an that they are also in involution, so
that, by Liouville's theorem, the system turns out to be integrable.
The definition of normal form given in~\frmref{nrm.2} is more
general, since it includes also the case of resonant frequencies.

The calculation of the normal form may be performed using the Lie
transform method, which turns out to be quite effective.  We give here
the algorithm without proof.  A complete description may be found,
e.g., in~\dbiref{Giorgilli-1978}, and the description of a program
implementing the method via computer algebra is given
in~\dbiref{Giorgilli-1979}.  The corresponding {\tt FORTRAN} program
is available from the CPC library.

The Lie
transform is defined as follows.  Let a {\corsivo generating sequence}
$\chi_1(x,y)$, $\chi_2(x,y),\ldots$ be given, and define the operator
$$
T_{\chi} = \sum_{s\ge 0} E_s
\formula{nrm.3}
$$
where the sequence $E_0,E_1,\ldots$ of operators is recursively
defined as
$$
E_0=1\ ,\quad
 E_s = \sum_{j=1}^{s}\frac{j}{s} L_{\chi_j}E_{s-j}
\formula{nrm.3a}
$$
This is a linear operator that is invertible and satisfies the
interesting properties
$$
T_{\chi} \{ f,g\} = \{T_{\chi}f,T_{\chi}g\}\ ,\quad
 T_{\chi}  (f\cdot g) = T_{\chi}f\cdot T_{\chi}g\ .
\formula{nrm.4}
$$
Let now $Z(x,y) = H_0(x,y) + Z_1(x,y) + Z_2(x,y)+\ldots\ $ be a
function such that
$$
T_{\chi} Z = H\ ,
\formula{nrm.5}
$$
where $H$ is our original Hamiltonian, and let $Z$ possess a first
integral $\Phi$, i.e., a function satisfying $\{Z,\Phi\}=0$.  Then one
has also
$$
T_{\chi}\{Z,\Phi\} = \{T_{\chi}Z,T_{\chi}\Phi\} = \{H,T_{\chi}\Phi\} =0\ ,
$$
which means that {\corsivo if $\Phi$ is a first integral for $Z$ then
$T_{\chi}\Phi$ is a first integral for $H$.}

The question now is: {\corsivo can we find a generating sequence
$\chi_1,\,\chi_2,\ldots$ such that the function $Z$
satisfying~\frmref{nrm.5} is in Birkhoff normal form?}

The answer to this question is in the positive, and the generating sequence
may be calculated via an explicit algorithm that can be effectively
implemented via computer algebra.  We include here the algorithm,
referring to, e.g.,~\dbiref{Giorgilli-1978} for a complete deduction.
Here we want only to stress that all operations that are required may
be actually implemented on a computer.

The generating sequence is determined by solving for
$\chi$ and $Z$ the equations
$$
Z_s - L_{H_0}\chi_s  = H_s + Q_s\ ,\quad
 s \ge 1\ ,
\formula{nrm.6}
$$
where $Q_s$ is a known homogeneous polynomial of degree $s+2$ given by
$Q_1=0$ and
$$
Q_s = -\sum_{j=1}^{s-1}
        \bigl( E_j Z_{s-j} + \frac{j}{s}\{\chi_j, E_{s-j}H_0\}\bigr)
\ ,\quad s\gt 1\ .
$$
In order to solve~\frmref{nrm.6} it is convenient to introduce
complex variables $\xi,\eta$ via the canonical transformation
$$
x_j = \frac{1}{\sqrt{2}} (\xi_j+i\eta_j)\ ,\quad
 y_j = \frac{i}{\sqrt{2}} (\xi_j-i\eta_j)
$$
which transforms $H_0 = i\sum_{j}\omega_j\xi_j\eta_j$.  In these
variables the operator $L_{H_0}$ takes a diagonal form, since
$$
L_{H_0} \xi^j\eta^k = i\langle k-j,\omega\rangle \xi^j\eta^k\ ,
$$
where we have used the multi-index notation
$\xi^j=\xi_1^{j_1}\cdot\ldots\cdot\xi_n^{j_n}$, and similarly for
$\eta$.  Thus, writing the r.h.s.\ of~\frmref{nrm.6} as a sum of
monomials $c_{j,k}\xi^j\eta^k$ the most direct form of the solution is
found by including in $Z$ all monomials with $\langle
k-j,\omega\rangle=0$, and adding $\frac{c_{j,k}}{i\langle
k-j,\omega\rangle}\xi^j\eta^k$ to $\chi_s$ for all monomials with
$\langle k-j,\omega\rangle\ne 0$.  This is the usual way of
constructing a normal form for the system~\frmref{nrm.1}.

Let us now examine in some more detail the algebraic aspect.  With
some patience one can verify that~\frmref{nrm.6} involves only
homogeneous polynomials of degree $s+2$.  Thus, one should be able to
manipulate this kind of functions.  Moreover, a careful examination of
the algorithm shows that there are just elementary algebraic
operations that are required, namely:

\item{(i)}sums and multiplication by scalar quantities;
\item{(ii)}Poisson brackets, which actually require derivatives of
monomials, sums and products;
\item{(iii)}linear substitution of variables, which may still be reduced to
calculation of sums and products without affecting the degree of the polynomial;
\item{(iv)}solving equation~\frmref{nrm.6}, which just requires a
division of coefficients.

\noindent
These remarks should convince the reader that implementing the
calculation of the normal form via algebraic manipulation on a
computer is just matter of being able of representing homogeneous
polynomials in many variables and performing on them a few elementary
operations, such as sum, product and derivative.

\subsection{1a.2}{A few elementary considerations}
In order to have an even better understanding the reader may want to
consider the elementary problem of representing polynomials in one
single variable.  We usually write such a polynomial of degree $s$
(non homogeneous, in this case) as
$$
f(x) = a_0 + a_1 x + \ldots + a_s x^s\ .
$$
A machine representation is easily implemented by storing the
coefficients $a_0$, $a_1,\ldots$,$\,a_n$ as a one-dimensional array of
floating point quantities, either real or complex.  E.g., in {\tt
FORTRAN} language one can represent a polynomial of degree 100 by just
saying, e.g., {\tt DIMENSION F(101)} and storing the coefficient $a_j$
as {\tt F(j+1)} (here we do not use the extension of {\tt FORTRAN}
that allows using zero or even negative indices for an array).
Similarly in a language like {\tt C} one just says, e.g., {\tt double
f[101]} and stores $a_j$ as {\tt f[j]}.

The operation of sum is a very elementary one: if $f,\,g$ are two
polynomials and the coefficients are stored in the arrays {\tt f,g}
(in {\tt C} language) then the sum $h$ is the array {\tt h} with elements
{\tt h[j] = f[j] + g[j]}.  The derivative of $f$ is the array {\tt fp}
with elements {\tt fp[j] = (j+1)*f[j+1]}.  In a similar way one can
calculate the product, by just translating in a programming language the
operations that are usually performed by hand.

The case of polynomials in two variables is just a bit more difficult.
A homogeneous polynomial of degree $s$ is usually written as
$$
f(x,y) = a_{s,0}x^s + a_{s-1,1}x^sy +\ldots+a_{0,s}y^s\ .
$$ 
The naive (not recommended) representation would use an array
with two indices (a matrix), by saying, e.g., {\tt DIMENSION
F(101,101)} and storing the coefficient $a_{j,k}$ as {\tt F(j+1,k+1)}.
Then the algebra is just a straightforward modification with respect
to the one-dimensional case.

Such a representation is not recommended for at least two reasons.
The first one is that arrays with arbitrary dimension are difficult to
use, or even not allowed, in programming languages.  The second and
more conclusive reason is that such a method turns out to be very
effective in wasting memory space. E.g., in the two dimensional case a
polynomial of degree up to $s$ requires a matrix with $(s+1)^2$
elements, while only $(s+1)(s+2)/2$ are actually used.  Things go
much worse in higher dimension, as one easily realizes.

The arguments above should have convinced the reader that an effective
method of representing polynomials is a basic tool in order to perform
computer algebra for problems like the calculation of normal form.
Once such a method is available, the rest is essentially known algebra,
that needs to be translated in a computer language.

The problem for Poisson series is a similar one, as the reader can
easily imagine.  The following sections contains a detailed discussion
of indexing methods particularly devised for polynomials and for
Poisson series.  The underlying idea is to represent the coefficients
as a one-dimensional array by suitably packing them in an effective
manner, so as to avoid wasting of space.

\section{2}{General scheme}
The aim of this section is to illustrate how an appropriate algebraic
structure may help in representing the particular classes of functions
that usually appear in perturbation theory.  We shall concentrate our
attention only on polynomials and trigonometric polynomials, which are
the simplest and most common cases.  However, the reader will see that
most of the arguments used here apply also to more general cases.

\subsection{funrep.1.1}{Polynomials and power series}
Let $\Pscr$ denote the vector space of polynomials in the independent
variables $x=(x_1,\ldots,x_n)\in\reali^{n}$.  A basis for this
vector space is the set $\{u_k(x)\}_{k\in\interi^n_+}$, where
$$
u_k(x)=x^k\equiv x_1^{k_1}\cdot\ldots\cdot x_n^{k_n}\ .
\formula{funrep.1}
$$
In particular, we shall consider the subspaces $\Pscr_s$ of $\Pscr$ that
contain all homogeneous polynomials of a given degree $s\ge 0$; the subspace
$\Pscr_0$ is the one-dimensional space of constants, and its basis is
$\{1\}$.  The relevant algebraic properties are the following:

\item{(i)}every subspace $\Pscr_s$ is closed with respect to 
sum and multiplication by a number, i.e., if $f\in\Pscr_s\ \wedge\
g\in\Pscr_s$ then $f+g\in\Pscr_s$ and $\alpha f\in\Pscr_s$;

\item{(ii)}the product of homogeneous polynomials is a
homogeneous polynomial, i.e., if $f\in\Pscr_r\ \wedge\ g\in\Pscr_s$ then 
$fg\in\Pscr_{r+s}$;

\item{(iii)}the derivative with respect to one variable maps
homogeneous polynomials into homogeneous polynomials, i.e., if
$f\in\Pscr_s$ then $\partial_{x_j}f\in\Pscr_{s-1}$; if $s=0$ then
$\partial_{x_j}f=0$, of course.  

\noindent These three properties are the basis for most of the algebraic
manipulations that are commonly used in perturbation theory.  

A power series is represented as a sum of homogeneous
polynomials.  Of course, in practical calculations the series will be
truncated at some order.  Since every homogeneous polynomial
$f\in\Pscr_s$ can be represented as
$$
f(x) = \sum_{|k|=s} f_k u_k(x)\ ,
$$
it is enough to store in a suitable manner the coefficients $f_k$.  A convenient way,
particularly effective when most of the coefficients are different from zero, is based on
the usual lexicographic ordering of polynomials (to be pedantic, inverse
lexicographic).  E.g., a homogeneous polynomial of degree $s$ in two variables is ordered
as
$$
a_{s,0} x_1^s + a_{s-1,1}x_1^{s-1}x_2 + \ldots + a_{0,s}x_2^s\ .
$$
The idea is to use the position of a monomial $x^{k}$ in the
lexicographic order as an index $I(k_1,\ldots,k_n)$ in an array of
coefficients.  We call $I$ and {\corsivo indexing function}.   Here we
illustrate how to use it, deferring to sect.~\secref{indexf.1} the
actual construction of the function.

\table{funrep.1}{Illustrating the function representation for power
series.  A memory block is assigned to the function $f(x)$.  The
coefficient $f_k$ of $u_k(x)$ is stored at the address resulting by
adding the offset $I(k)$ to the starting address of the memory block.}{ 
   \begingroup
\ungr=.014pt   
\smspes=0.5pt
\def\oofst{1000}
\def\vofst{0}
{\hbox{\parindent 0pt
\initblk{7000}
\elm{1000}{$f_{(0,0,\ldots,0)}$}{$\leftarrow 0=I(k) \leftarrow k=(0,0,\ldots,0)$}
\elm{1000}{$f_{(1,0,\ldots,0)}$}{$\leftarrow 1=I(k) \leftarrow k=(1,0,\ldots,0)$}
\elmnol{1000}{\null}{\null}
\elmnol{500}{$\ldots$}{\null}
\elmnol{500}{$\ldots$}{\null}
\elmnol{500}{$\ldots$}{\null}
\elm{500}{\null}{\null}
\elm{1000}{$f_{(k_1,k_2,\ldots,k_n)}$}{$\leftarrow \>I(k)\> 
   \leftarrow k=(k_1,k_2,\ldots,k_n)$}
\elmnol{1000}{\null}{\null}
\elmnol{500}{$\ldots$}{\null}
\elmnol{500}{$\ldots$}{\null}
\elmnol{500}{$\ldots$}{\null}
\elm{500}{\null}{\null}
}}
\endgroup
 
  }

The method is illustrated in table~\tabref{funrep.1}.  Let $f$ be a
power series, truncated at some finite order $s$.  A memory block is
assigned to $f$.  The size of the block is easily determined as
$I\bigl((0,\ldots,0,s)\bigr)$.  For,
$(0,\ldots,0,s)$ is the last vector of length $s$.  The starting
address of the block is assigned to the coefficient of
$u_{(0,0,\ldots,0)}$; the next address is assigned to the coefficient
of $u_{(1,0,\ldots,0)}$, because $(1,0,\ldots,0)$ is the first vector
of length $1$, and so on.  Therefore, the address assigned to the
coefficient of $u_{(k_1,\ldots,k_n)}$ is the starting address of the
block incremented by $I\bigl((k_1,\ldots,k_n)\bigr)$.  If $f$ is a
homogeneous polynomial of degree $s$ the same scheme works fine with a
few minor differences: the length of the block is
$I\bigl((0,\ldots,0,s)\bigr)-I\bigl((0,\ldots,0,s-1)\bigr)$, the
starting address of the block is associated to the coefficient of
$u_{(s,0,\ldots,0)}$, and the coefficient of $u_{(k_1,\ldots,k_n)}$ is
stored at the relative address
$I\bigl((k_1,\ldots,k_n)\bigr)-I\bigl((0,\ldots,0,s-1)\bigr)$.  This
avoids leaving an empty space at the top of the memory block.

In view of the form above of the representation a function is
identified with a set of pairs $(k,f_k)$, where $k\in\interi_+^n$ is
the vector of the exponents, acting as the label of the elements of
the basis, and $f_k$ is the numerical coefficient.  Actually the
vector $k$ is not stored, since it is found via the index. The
algebraic operations of sum, product and differentiation can be
considered as operations on the latter set.

\item{(i)}If $f,g\in\Pscr_s$ then the operation of calculating
the sum $f+g$ is represented as

$$
\left.
\vcenter{\openup1\jot\halign{
\hfil$\displaystyle({#},$
&$\displaystyle{#})$\hfil
\cr
k & f_k\cr
k & g_k\cr
}}
\right\}\mapsto
(k,f_k+g_k)\ ,
$$

\item{}to be executed over all $k$ such that $|k|=s$.

\item{(ii)}If $f\in\Pscr_r$ and $g\in\Pscr_s$ then the operation
of calculating the product $fg$ is represented as

$$
\left.
\vcenter{\openup1\jot\halign{
\hfil$\displaystyle({#},$
&$\displaystyle{#})$\hfil
\cr
k & f_k\cr
k' & g_{k'}\cr
}}
\right\}\mapsto
(k+k',f_kg_{k'})\ ,
$$

\item{}to be executed over all $k,k'$ such that $|k|=r$ and $|k'|=s$.

\item{(iii)}If $f\in\Pscr_s$ then the operation of differentiating
$f$ with respect to, e.g., $x_1$ is represented as

$$
(k,f_k) \mapsto
\left\{
\vcenter{\openup1\jot\halign{
$\displaystyle{#}$\hfil
&\quad{\rm for\ }$\displaystyle{#}$\hfil
\cr
\emptyset & k_1=0\ ,\cr
(k',k_1 f_k) & k_1\ne 0\ ,
\cr
}}
\right.
$$

\item{}where $k'=(k_1-1,k_2,\ldots,k_n)$.

It is perhaps worthwhile to spend a few words about how to make the
vector $k$ to run over all its allowed values.  In the case of sum, we
do not really need it: since the indexes of both addends and of the
result are the same, the operation can actually be performed no matter
which $k$ is involved: just check that the indexes are in the correct
range.\footnote{For a homogeneous polynomial of degree $s$ the first
vector is $(s,0,\ldots,0)$, and the last one is $(0,\ldots,0,s)$.  The
indexes of these two vectors are the limits of the indexes in the
sum.} In order to perform product and differentiation it is essential
to know the values of $k$ and $k'$.  To this end, we can either use the
inverse of the indexing function, or generate the whole sequence by
using a function that gives the vector next to a given $k$.

\subsection{funrep.1.2}{Fourier series}
Let us denote by $\phi=(\phi_1\ldots,\phi_n)\in\toro^n$ the
independent variables.  The Fourier expansion of a real function on
$\toro^n$ takes the form
$$
f(\phi)=\sum_{k\in\interi^n}\left(a_k\cos\langle k,\phi\rangle
+b_k\sin\langle k,\phi\rangle\right)\ ,
\formula{funrep.5}
$$
where $a_k$ and $b_k$ are numerical coefficients.  In this
representation there is actually a lot of redundancy: in view of
$\cos(-\alpha)=\cos\alpha$ and $\sin(-\alpha)=-\sin\alpha$ the modes
$-k$ and $k$ can be arbitrarily interchanged.  On the other hand, it
seems that we actually need two different arrays for the sin and cos
components, respectively.  A straightforward way out is to use the
exponential representation $\sum_{k} a_k e^{i\langle k,\phi\rangle}$,
but a moment's thought leads us to the conclusion that the redundancy
is not removed at all.  However, we can at the same time remove the
redundancy and reduce the representation to a single array by
introducing a suitable basis $\{u_k(\phi)\}_{k\in\interi^n}\,$.  Let
$k\in\interi^n$; we shall say that $k$ is {\corsivo even} if the first
non zero component of $k$ is positive, and that $k$ is {\corsivo odd}
if the first non zero component of $k$ is negative.  The null vector
$k=0$ is said to be even.  Then we set
$$
u_k(\phi)=\left\{
\vcenter{\openup1\jot\halign{
$\displaystyle{#}$\hfil
&\quad{\rm for\ }$\displaystyle{#}$\hfil
\cr
\cos\langle k,\phi\rangle & k{\ \rm even}\ ,\cr
\sin\langle k,\phi\rangle & k{\ \rm odd}\ .\cr
}}
\right.
\formula{funrep.6}
$$
This makes the representation $f(\phi)=\sum_{k\in\interi^n}\phi_k
u_{k}(\phi)$ unique and redundancy free.  It may be convenient to
remark that the notation for the sin function may create some
confusion.  Usually, working with one variable, we write $\sin\phi$.
The convention above means that we should rather write $-\sin(-\phi)$,
which is correct, but a bit funny.  This should be taken into account
when, after having accurately programmed all the operations, we
discover that our manipulator says, e.g., that $\der{}{\phi}\cos\phi =
-\sin(-\phi)$.

In view of the discussion in the previous section it should now be
evident that a truncated Fourier expansion of a function $f(\phi)$ can
easily be represented by storing the coefficient of $u_{k}(\phi)$ at
an appropriate memory address, as calculated by the indexing function
$I(k)$ of sect.~\secref{indexf.2}.

The considerations of the previous section can be easily extended to
the problem of calculating the sum and/or product of two functions,
and of differentiating a function.  Let us identify any term of the
Fourier expansion of the function $f$ with the pair $(k,f_k)$.  Let us
also introduce the functions $\odd(k)$ and $\even(k)$ as follows: if
$k$ is odd, then $\odd(k)=k$ and $\even(k)=-k$; else $\odd(k)=-k$ and
$\even(k)=k$.  That, is, force $k$ to be odd or even, as needed, by
possibly changing its sign.

\item{(i)}Denoting by $(k,f_k)$ and $(k,g_k)$ the same
Fourier components of two functions $f$ and $g$, respectively, the sum
is computed as

$$
\left.
\vcenter{\openup1\jot\halign{
\hfil$\displaystyle({#},$
&$\displaystyle{#})$\hfil
\cr
k & f_k\cr
k & g_k\cr
}}
\right\}\mapsto
(k,f_k+g_k)\ .
\formula{funrep.7}
$$

\item{(ii)}Denoting by $(k,f_k)$ and $(k',g_{k'})$ any two terms
in the Fourier expansion of the functions $f$ and $g$, respectively,
the product is computed as

$$
\left.
\vcenter{\openup1\jot\halign{
\hfil$\displaystyle({#},$
&$\displaystyle{#})$\hfil
\cr
k & f_k\cr
k' & g_{k'}\cr
}}
\right\}\mapsto
\left\{
\vcenter{\openup1\jot\halign{
\hfil$\displaystyle{#}$
&$\displaystyle{#}$\hfil
&\ {\rm for}$\>\displaystyle{#}$\hfil
\cr
\left(\even(k+k'),\frac{f_kg_{k'}}{2}\right)
 & \cup \left(\even(k-k'),\frac{f_kg_{k'}}{2}\right)
  & k\>{\rm even}\,,\>k'\>{\rm even}\>,
\cr
\left(\odd(k+k'),\frac{f_kg_{k'}}{2}\right)
 & \cup \left(\odd(k-k'),-\frac{f_kg_{k'}}{2}\right)
  & k\>{\rm even}\,,\>k'\>{\rm odd}\>,
\cr
\left(\odd(k+k'),\frac{f_kg_{k'}}{2}\right)
 & \cup \left(\odd(k-k'),\frac{f_kg_{k'}}{2}\right)
  & k\>{\rm odd}\,,k'\>{\rm even}\>,
\cr
\left(\even(k+k'),-\frac{f_kg_{k'}}{2}\right)
 & \cup \left(\even(k-k'),\frac{f_kg_{k'}}{2}\right)
  & k\>{\rm odd}\,,\>k'\>{\rm odd}\>.
\cr
}}
\right.
\formula{funrep.8}
$$
Remark that the product always produces two distinct terms, unless
$k=0$ or $k'=0$.

\item{(iii)}Denoting by $(k,f_k)$ any term in the Fourier
expansion of a function $f$, differentiation with respect to, e.g.,
$\phi_1$ is performed as

$$
(k,f_k) \mapsto
\left\{
\vcenter{\openup1\jot\halign{
$\displaystyle{#}$\hfil
&\quad{\rm for\ }$\displaystyle{#}$\hfil
\cr
(-k,-k_1 f_k) & k{\ \rm even}\ ,\cr
(-k, k_1 f_k) & k{\ \rm odd}\ .\cr
}}
\right.
\formula{funrep.9}
$$
All these formul{\ae} follow from well known trigonometric
identities.

\section{3}{Indexing functions}
The basic remark for constructing an index function is the
following.   Suppose that we are given a countable set $\Ascr$.  Suppose
also that $\Ascr$ is equipped with a relation of complete ordering,
that we shall denote by the symbols $\prec$, $\preceq$, $\succ$ and $\succeq$.   So,
for any two elements $a,b\in\Ascr$ exactly one of the relations
$a\prec b$, $a=b$ and $b\succ a$ is true.   Suppose also that there is a minimal
element in $\Ascr$, i.e., there is $a_0\in\Ascr$ such that $a\succ a_0$
for all $a\in\Ascr$ such that $a\ne a_0$.   Then an index function
$I$ is naturally defined as
$$
I(a)=\#\{b\in\Ascr\>:\>b\prec a\}\ .   
\formula{indexf.0}
$$ 
If $\Ascr$ is a finite set containing $N$ elements, then
$I(\Ascr)=\{0,1,\ldots,N-1\}$.  If $\Ascr$ is an infinite (but
countable) set, then $I(\Ascr)=\interi_+$, the set of non negative
integers.  For instance, the trivial case is $\Ascr=\interi_+$ equipped
with the usual ordering relation.  In such a case the indexing function
is just the identity.

Having defined the function $I(a)$, we are interested in performing
the following basic operations:

\item{(i)}for a given $a\in\Ascr$, find the index $I(a)$;

\item{(ii)}for a given $a\in\Ascr$, find the element next 
(or prior) to $a$, if it exists;

\item{(iii)}for a given $l\in I(\Ascr)$, find $I^{-1}(l)$, i.e.,
the element $a\in\Ascr$ such that $I(a)=l$.

The problem here is to implement an effective construction of the
index  for some particular subsets of $\interi^{n}$ that
we are interested in.   In order to avoid confusions, we shall use the
symbols $\prec$, $\preceq$, $\succ$ and $\succeq$ when dealing with an
ordering relation in the subset of $\interi^{n}$ under consideration.
The symbols $\lt$, $\le$, $\ge$ and $\gt$ will always denote the
usual ordering relation between integers.

As a first elementary example, let us consider the case
$\Ascr=\interi$.   The usual ordering relation $\lt$ does not fulfill
our requests, because there is no minimal element.  However, we can
construct a different ordering satisfying our requests as follows.

\blankq
\begingroup
\noindent\enufnt
Let $k,k'\in\interi$.  We shall say that $k'\prec k$ in case one of the
following relations is true:
\item{(i)}$|k'|\lt |k|\>${\rm;}
\item{(ii)}$|k'|=|k|\ \wedge\ k'\gt k\>$.

\endgroup

\blankq\noindent
The resulting order is $0,1,-1,2,-2,\ldots\,$, so that $0$ is the
minimal element.

Constructing the indexing function in this case is easy.   Indeed, we have
$$
I(0)=0\ ,\quad
I(a)=\left\{
\vcenter{\openup1\jot\halign{
$\displaystyle{#}$\hfil
&\quad{\rm for\ }$\displaystyle{#}$\hfil
\cr
2a-1&a\gt 0\ ,
\cr
-2a&a\lt 0\ .
\cr
}}
\right.
\formula{indexf.0.1}
$$
The inverse function is also easily constructed:
$$
I^{-1}(0)=0\ ,\quad
I^{-1}(l)=
\left\{
\vcenter{\openup1\jot\halign{
$\displaystyle{#}$\hfil
&\quad{\rm for\ }$\displaystyle{#}$\hfil
\cr
(l+1)/2& l\ {\rm odd}\ ,
\cr
-l/2&l\ {\rm even}\ .
\cr
}}
\right.
\formula{indexf.0.2}
$$
In the rest of this section we show how an indexing function can be
constructed for two particularly interesting cases, namely polynomials
and trigonometric polynomials.   However, we stress that the procedure
we are using is a quite general one, so it can be extended to other
interesting situations.

\section{indexf.1}{The polynomial case}
Let us first take $\Ascr_n=\interi^n_+$, i.e., integer vectors
with non negative components; formally
$$
\Ascr_n=\{k=(k_1,\ldots,k_n)\in\interi^n\>:\>k_1\ge 0,\ldots,k_n\ge
0\}\ .
$$
The index $n$ in $\Ascr_n$ denotes the dimension of the space.  This
case is named ``polynomial'' because it occurs precisely in the
representation of homogeneous polynomials, and so also in the Taylor
expansion of a function of $n$ variables: the integer vectors
$k\in\Ascr_n$ represent all possible exponents.

We shall denote by $|k|=k_1+\ldots+k_n$ the length (or norm) of the
vector $k\in\interi^n_+$.  Furthermore, to a given vector
$k=(k_1,\ldots,k_n)\in\Ascr_n$ we shall associate the vector
$t(k)\in\Ascr_{n-1}$ (the {\corsivo tail} of $k$) defined as
$t(k)=(k_2,\ldots,k_n)$.  This definition is meaningful only if $n\gt
1$, of course.

\subsection{indexf.1.1}{Ordering relation}
Pick a fixed $n$, and consider the finite family of sets
$\Ascr_{1}=\interi_+,\ldots,\Ascr_{n}=\interi^n_+$.

\blankq
\begingroup
\noindent\enufnt
Let $k,k'\in\Ascr_m$, with any $1\le m\le n$.  We shall say that
$k'\prec k$ in case one of the following conditions is true:

\item{(i)}$m\ge 1\ \wedge\ |k'|\lt |k|\>${\rm ;}
\item{(ii)}$m\gt 1\ \wedge\ |k'|=|k|\ \wedge\ k'_1\gt k_1\>${\rm ;}
\item{(iii)}$m\gt 1\ \wedge\ |k'|=|k|\ \wedge\ k'_1=k_1
\ \wedge\ t(k')\prec t(k)\>$.  

\endgroup

\table{indexf.1}{Ordering of integer vectors in $\interi^m_+$ 
for $m=2,3,4,5$.}  
{ 
   \begingroup
$$
\vcenter{\openup1\jot\halign{
\quad\hfil$\displaystyle{#}$\quad
&\quad\hfil$\displaystyle{#}$\hfil\quad
&\quad\hfil$\displaystyle{#}$\hfil\quad
&\quad\hfil$\displaystyle{#}$\hfil\quad
&\quad\hfil$\displaystyle{#}$\hfil\quad
\cr
I(k)    &  m=2   &  m=3     &  m=4       &  m=5         \cr
\noalign{\vskip 2pt}
\noalign{\hrule}
\noalign{\vskip 3pt}
0\quad  &  (0,0) &  (0,0,0) &  (0,0,0,0) &  (0,0,0,0,0) \cr
1\quad  &  (1,0) &  (1,0,0) &  (1,0,0,0) &  (1,0,0,0,0) \cr
2\quad  &  (0,1) &  (0,1,0) &  (0,1,0,0) &  (0,1,0,0,0) \cr
3\quad  &  (2,0) &  (0,0,1) &  (0,0,1,0) &  (0,0,1,0,0) \cr
4\quad  &  (1,1) &  (2,0,0) &  (0,0,0,1) &  (0,0,0,1,0) \cr
5\quad  &  (0,2) &  (1,1,0) &  (2,0,0,0) &  (0,0,0,0,1) \cr
6\quad  &  (3,0) &  (1,0,1) &  (1,1,0,0) &  (2,0,0,0,0) \cr
7\quad  &  (2,1) &  (0,2,0) &  (1,0,1,0) &  (1,1,0,0,0) \cr
8\quad  &  (1,2) &  (0,1,1) &  (1,0,0,1) &  (1,0,1,0,0) \cr
9\quad  &  (0,3) &  (0,0,2) &  (0,2,0,0) &  (1,0,0,1,0) \cr
10\quad &  (4,0) &  (3,0,0) &  (0,1,1,0) &  (1,0,0,0,1) \cr
11\quad &  (3,1) &  (2,1,0) &  (0,1,0,1) &  (0,2,0,0,0) \cr
12\quad &  (2,2) &  (2,0,1) &  (0,0,2,0) &  (0,1,1,0,0) \cr
13\quad &  (1,3) &  (1,2,0) &  (0,0,1,1) &  (0,1,0,1,0) \cr
14\quad &  (0,4) &  (1,1,1) &  (0,0,0,2) &  (0,1,0,0,1) \cr
15\quad &  (5,0) &  (1,0,2) &  (3,0,0,0) &  (0,0,2,0,0) \cr
16\quad &  (4,1) &  (0,3,0) &  (2,1,0,0) &  (0,0,1,1,0) \cr
17\quad &  (3,2) &  (0,2,1) &  (2,0,1,0) &  (0,0,1,0,1) \cr
18\quad &  (2,3) &  (0,1,2) &  (2,0,0,1) &  (0,0,0,2,0) \cr
19\quad &  (1,4) &  (0,0,3) &  (1,2,0,0) &  (0,0,0,1,1) \cr
20\quad &  (0,5) &  (4,0,0) &  (1,1,1,0) &  (0,0,0,0,2) \cr
\cdots\quad&\cdots& \cdots  &  \cdots    &  \cdots      \cr
\noalign{\vskip 2pt}
\noalign{\hrule}
}}
$$
\endgroup
  
}

\blankq\noindent
In table~\tabref{indexf.1} the ordering resulting from this definition
is illustrated for the cases $m=2,3,4,5$.

\blankq If $m=1$ then only (i) applies, and this ordering coincides
with the natural one in $\interi_+$.  For $m\gt 1$, if (i) and (ii) are
both false, then (iii) means that one must decrease the dimension $n$
by replacing $k$ with its tail $t(k)$, and retry the comparison.  For
this reason the ordering has been established for $1\le m\le
n$.  Eventually, one ends up with $m=1$, to which only (i) applies.

It is convenient to define 
$\Pscr_n(k)$ as the set of the elements which precede $k$; formally:
$$
\Pscr_n(k)=\{k'\in\Ascr_n\>:\>k'\prec k\}\ .
$$
With the latter notation the indexing function is simply defined as
$I(k)=\#\Pscr_n(k)$.  The following definitions are also useful.   Pick
a vector $k\in\Ascr_n$, and define the sets $\Bscr_n^{(i)}(k)$,
$\Bscr_n^{(ii)}(k)$ and $\Bscr_n^{(iii)}(k)$ as the subsets of
$\Ascr_n$ satisfying (i), (ii) and (iii), respectively, in the
ordering algorithm above.  Formally:
$$
\eqalign{ \Bscr_n^{(i)}(0) &=\Bscr_n^{(ii)}(0)=\Bscr_n^{(iii)}(0)
 =\Bscr_1^{(ii)}(k)=\Bscr_1^{(iii)}(k)=\emptyset\ ,
\cr
\Bscr_n^{(i)}(k)
&=\{k'\in\Ascr_n\>:\>|k'|\lt |k|\}\ ,
\cr
\Bscr_n^{(ii)}(k)
&=\{k'\in\Ascr_n\>:\>|k'|=|k|\ \wedge\ k'_1\gt k_1\}\ ,
\cr
\Bscr_n^{(iii)}(k)
&=\{k'\in\Ascr_n\>:\>|k'|=|k|\ \wedge\ k'_1=k_1
\ \wedge\ t(k')\lt t(k)\}\ .
\cr
}
\formula{indexf.7}
$$
The sets $\Bscr_n^{(i)}(k)$, $\Bscr_n^{(ii)}(k)$ and
$\Bscr_n^{(iii)}(k)$ are pairwise disjoint, and moreover
$$
\Bscr_n^{(i)}(k)\cup \Bscr_n^{(ii)}(k) \cup \Bscr_n^{(iii)}(k)
=\Pscr_n(k)\ .
$$
This easily follows from the definition.  

\subsection{indexf.1.2}{Indexing function for polynomials}
Let $k\in\Ascr_n$.  In view of the definitions and of the properties
above the index function, defined as in~\frmref{indexf.0}, turns out
to be
$$
I(0)=0\ ,\quad
I(k)= \#\Bscr_n^{(i)}(k) +\#\Bscr_n^{(ii)}(k) 
      +\#\Bscr_n^{(iii)}(k)\ .
\formula{indexf.8}
$$

Let us introduce the functions
$$
\eqalign{
J(n,s)&=\#\{k\in\Ascr_n\>:\>|k|=s\}\ ,\cr
N(n,s)&=\sum_{j=0}^{s}J(n,j)\qquad\qquad\qquad{\rm for\ }n\ge
1\,,\>s\ge 0\ .\cr
}
\formula{indexf.10}
$$
These functions will be referred to in the following as {\corsivo
$J$-table and $N$-table}.

We claim that the indexing function can be recursively computed as
$$
\eqalign{
I(0)&=0\ ,
\cr
I(k)&=\left\{
\vcenter{\openup1\jot\halign{
$\displaystyle{#}$\hfil
&\quad{\rm for\ }$\displaystyle{#}$\hfil
\cr
N(n,|k|-1) & k_1=|k|\ ,
\cr
N(n,|k|-1)+I(t(k)) & k_1\lt|k|\ .
\cr
}}\right.
\cr
}
\formula{indexf.100}
$$
The claim follows from 
$$
\formdef{indexf.20}
\formdef{indexf.21}
\formdef{indexf.22}
\leqalignno{
\#\Bscr_n^{(i)}(k) &= N(n,|k|-1)\ ;
&\frmref{indexf.20}
\cr
\#\Bscr_n^{(ii)}(k) 
&=\left\{\vcenter{\openup1\jot\halign{
\hbox to 13pc{$\displaystyle{#}$\hfil}\hfil
&\quad{\rm for\ }$\displaystyle{#}$\hfil
\cr
0 & k_1=|k|\ ,
\cr
N(n-1,|k|-k_1-1) & k_1\lt |k|\ ;
\cr
}}\right.
&\frmref{indexf.21}
\cr
\#\Bscr_n^{(iii)}(k)
&=\left\{\vcenter{\openup1\jot\halign{
\hbox to 13pc{$\displaystyle{#}$\hfil}\hfil
&\quad{\rm for\ }$\displaystyle{#}$\hfil
\cr
I(t(k)) & k_1=|k|\ ,
\cr
I(t(k))-N(n-1,|k|-k_1-1) & k_1\lt |k|\ .
\cr
}}\right.
&\frmref{indexf.22}
\cr
}
$$
The equality~\frmref{indexf.20} is a straightforward consequence of
the definition of the $N$-table.  The equality~\frmref{indexf.21}
follows from~\frmref{indexf.7}.  Indeed, for $|k|=k_1$ we have
$\Bscr_n^{(ii)}(k)=\emptyset$, and for $|k|\gt k_1$ we
have
$$
\eqalign{
\Bscr_n^{(ii)}(k)
&= \bigcup_{k_1\lt j\le |k|} 
    \{k'\in\Ascr_n\>:\>k'_1=j\ \wedge\ |t(k')|=|k|-j\}
\cr
&= \bigcup_{0\le l\lt |k|-k_1} 
    \{k'\in\Ascr_n\>:\>k'_1=|k|-l\ \wedge\ |t(k')|=l\}\ .
\cr
}
$$
Coming to~\frmref{indexf.22}, first remark that
$$
\Bscr_n^{(iii)}(k)
=\{k'\in\Ascr_n\>:\>k'_1=k_1\ \wedge\ |t(k')|=|k|-k_1
  \ \wedge\ t(k')\prec t(k)\}\ ,
$$
so that 
$$
\#\Bscr_n^{(iii)}(k)=\#\{\lambda\in\Ascr_{n-1}\>:\>|\lambda|=|k|-k_1
\ \wedge\ \lambda\prec t(k)\}\ .
$$
Then, the equality follows by remarking that 
$$
\Pscr_{n-1}(t(k))=
 \{\lambda\in\Ascr_{n-1}\>:\>|\lambda|=|k|-k_1
  \ \wedge\ \lambda\prec t(k)\}
   \cup
    \{\lambda\in\Ascr_{n-1}\>:\>|\lambda|\lt |k|-k_1\}\ .
$$
Adding up all contributions~\frmref{indexf.100} follows.

\subsection{indexf.1.3}{Construction of the tables}
In view of~\frmref{indexf.10} and~\frmref{indexf.100} the indexing
function is completely determined in explicit form by the
$J$-table.  We show now how to compute the $J$-table recursively.  For
$n=1$ we have, trivially, $J(1,s)=1$ for $s\ge 0$.  For $n\gt 1$ use
the elementary property
$$
\{k\in\Ascr_n\>:\>|k|=s\}= 
 \bigcup_{0\le j\le s}\{k\in\Ascr_n\>:\>k_1=s-j
  \ \wedge\ |t(k)|=j\}\ .
$$
Therefore
$$
\eqalign{
J(1,s)&=1\ ,
\cr
J(n,s)&=\sum_{j=0}^{s}J(n-1,j)\quad {\rm for\ }n\gt 1\ .
\cr
}
\formula{indexf.11}
$$
This also means that, according to~\frmref{indexf.10}, we have
$N(n,s)=J(n+1,s)\,$.

By the way, one will recognize that the $J$-table is actually the
table of binomial coefficients, being
$J(n,s)=\left({n+s-1}\atop{n-1}\right)\,$.

\subsection{indexf.1.4}{Inversion of the index function}
The problem is to find the vector $k\in\interi_+^n$ 
corresponding to a given index $l$.  

For $n=1$ we have $I^{-1}(l)=l$, of course.  Therefore, let us assume
$n>1$.  We shall construct a recursive algorithm which calculates the
inverse function by just showing how to determine $k_1$ and
$I\bigl(t(k)\bigr)$.  

\item{(i)}If $l=0$, then $k=0$, and there is nothing else to do.

\item{(ii)}If $l>0$, find an integer $s$ satisfying $N(n,s-1)\le l
\lt N(n,s)$.  In view of \frmref{indexf.100} we have $|k|=s$ and
$I(t(k))=l-N(n,s-1)$.  Hence, by the same method, we can determine
$\bigl|t(k)\bigr|$, and so also $k_1=s-\bigl|t(k)\bigr|$.

\def\sourcecode{\par\vskip 2pt plus 1pt\begingroup
               \noindent\hangindent\parindent\obeylines\tt}
\def\endsourcecode{\par\endgroup\vskip 2pt plus 1pt}

\subsection{indexf.5}{An example of implementation}
We include here an example of actual implementation of the indexing
scheme for polynomials.  This is part of a program for the calculation
of first integrals that is fully described
in~\dbiref{Giorgilli-1979}.  The complete computer code is also
available from the CPC program library.

We should mention that the {\tt FORTRAN} code included here has been
written in 1976.  Hence it may appear a little strange to programmers
who are familiar with the nowadays compilers, since it does not use
many features that are available in {\tt FORTRAN~90 } or in the
current versions of the compiler.  It rather uses the standard of {\tt
FORTRAN~II}, with the only exception of the statement {\tt PARAMETER}
that has been introduced later.

The {\tt PARAMETER}s included in the code allow the user to control
the allocation of memory, and may be changed in order to adapt the
program to different needs.

\noindent
{\tt NPMAX} is the maximum number of degrees of freedom  

\noindent
{\tt NORDMX} is the maximal polynomial degree that will be used

\noindent
{\tt NBN1} and {\tt NBN2} are calculated from the previous parameters,
and are used in order to allocate the correct amount of memory for the
table of binomial coefficients.  Here are the statements:

\sourcecode
~~~~~~~PARAMETER (NPMAX=3)
~~~~~~~PARAMETER (NORDMX=40)
~~~~~~~PARAMETER (NBN2=2*NPMAX)
~~~~~~~PARAMETER (NBN1=NORDMX+NBN2)
\endsourcecode

As explained in the previous sections, the indexing function for
polynomials uses the table of binomial coefficients.  The table is
stored in a common block named {\tt BINTAB} so that it is available to
all program modules.  In the same block there are also some constants
that are used by the indexing functions and are defined in the
subroutine {\tt BINOM} below.  Here is the statement that must be
included in every source module that uses these data:

\sourcecode
~~~~~~~COMMON~~/BINTAB/~~IBIN(NBN1,NBN2),NPIU1,NMEN1,NFAT,NBIN
\endsourcecode

Subroutine {\tt BINOM} fills the data in the common block {\tt
BINTAB}.  It must be called at the beginning of the execution, so that
the constants become available.  Forgetting this call will produce
unpredictable results.  The calling arguments are the following.

\noindent
{\tt NLIB}\ :\ the number of polynomial variables.  In the Hamiltonian
case considered in the present notes it must be set as $2n$, where $n$
is the number of degrees of freedom.  It must not exceed the value of
the parameter {\tt NPMAX}.

\noindent
{\tt NORD}\ :\ the wanted order of calculation of the polynomials,
which in our case is the maximal order of the normal form.  It must
not exceed the value of the parameter {\tt NFAT}.

The subroutine checks the limits on the calling arguments; if the
limits are violated then the execution is terminated with an error
message.  The calculation of the part of the table of binomial
coefficients that will be used is based on well known formul{\ae}.

\sourcecode
~~~~~~~SUBROUTINE~BINOM(NLIB,NORD)
C
C~~~~~~Compute the table of the binomial coefficients.
C
~~~~~~~COMMON~~/BINTAB/~~IBIN(NBN1,NBN2),NPIU1,NMEN1,NFAT,NBIN
C
~~~~~~~NFAT=NORD+NLIB
~~~~~~~NBIN=NLIB
~~~~~~~IF(NFAT.GT.NBN1.OR.NBIN.GT.NBN2) GO TO 10
~~~~~~~NPIU1 = NLIB+1
~~~~~~~NMEN1 = NLIB-1
~~~~~~~DO 1 I=1,NFAT
~~~~~~~IBIN(I,1) = I
~~~~~~~DO 1 K=2,NBIN
~~~~~~~IF(I-K) 2,3,4
~2~~~~~IBIN(I,K) = 0
~~~~~~~GO TO 1
~3~~~~~IBIN(I,K) = 1
~~~~~~~GO TO 1
~4~~~~~IBIN(I,K) = IBIN(I-1,K-1)+IBIN(I-1,K)
~1~~~~~CONTINUE
~~~~~~~RETURN
~10~~~~WRITE(6,1000) NFAT,NBIN
~~~~~~~STOP
~1000~~FORMAT(//,5X,15HERROR SUB BINOM,2I10,//)
~~~~~~~END
\endsourcecode

Function {\tt INDICE} implements the calculation of the indexing
function for polynomials.  The argument {\tt J} is an integer array of
dimension {\tt NLIB} which contains the exponents of the monomial.  It
must contain non negative values with sum not exceeding the value {\tt
NORD} initially passed to the subroutine {\tt BINOM}.  These limits
are not checked in order to avoid wasting time: note that this
function may be called several millions of times in a program.  The
code actually implements the recursive formula~\frmref{indexf.100}
using iteration.  Recall that recursion was not implemented in {\tt
FORTRAN~II}.

\sourcecode
~~~~~~~FUNCTION INDICE(J,NLIB)
C
C~~~~~~Compute the relative address I corresponding to the
C~~~~~~exponents J.
C
~~~~~~~COMMON~~/BINTAB/~~IBIN(NBN1,NBN2),NPIU1,NMEN1,NFAT,NBIN
~~~~~~~DIMENSION J(NLIB)
C
~~~~~~~NP=NLIB+1
~~~~~~~INDICE = J(NLIB)
~~~~~~~M = J(NLIB)-1
~~~~~~~DO 1 I=2,NLIB
~~~~~~~IB=NP-I
~~~~~~~M = M + J(IB)
~~~~~~~IB=M+I
~~~~~~~INDICE = INDICE + IBIN(IB,I)
~1~~~~~CONTINUE
~~~~~~~RETURN
~~~~~~~END
\endsourcecode

Subroutine {\tt ESPON} is the inverse of the indexing function.
Given the index {\tt N} it calculates the array {\tt J} of dimension
{\tt NLIB} which contains the exponents.  The value of {\tt N} must be
positive (not checked) and must not exceed the maximal index
implicitly introduced by the initial choice of {\tt NLIB} and {\tt
NORD} passed to {\tt BINOM}.  The latter error is actually checked
(this does not increase the computation time).  The code implements
the recursive algorithm described in sect.~\sbsref{indexf.1.4},
again using iteration.


\sourcecode
~~~~~~~SUBROUTINE ESPON(N,J,NLIB)
C
C~~~~~~Compute the exponents J correponding to the
C~~~~~~index N.
C
~~~~~~~COMMON~~/BINTAB/~~IBIN(NBN1,NBN2),NPIU1,NMEN1,NFAT,NBIN
~~~~~~~DIMENSION J(NLIB)
C
~~~~~~~NM=NLIB-1
~~~~~~~NP=NLIB+1
~~~~~~~DO 1 K=NLIB,NFAT
~~~~~~~IF (N.LT.IBIN(K,NLIB)) GO TO 2
~1~~~~~CONTINUE
~~~~~~~WRITE(6,1000)
~~~~~~~STOP
~2~~~~~NN = K-1
~~~~~~~M = N-IBIN(NN,NLIB)
~~~~~~~IF(NLIB-2) 8,6,7
\goodbreak                         
~7~~~~~DO 3 I = 2,NM
~~~~~~~L = NP-I
~~~~~~~DO 4 K=L,NFAT
~~~~~~~IF(M.LT.IBIN(K,L)) GO TO 5
~4~~~~~CONTINUE
~5~~~~~IB=NLIB-L
~~~~~~~J(IB) = NN-K
~~~~~~~NN = K-1
~~~~~~~M = M - IBIN(NN,L)
~3~~~~~CONTINUE
~6~~~~~J(NM) = NN-M-1
~~~~~~~J(NLIB) = M
~~~~~~~RETURN
~8~~~~~J(1)=N
~~~~~~~RETURN
~1000~~FORMAT(//,5X,15HERROR SUB ESPON,//)
~~~~~~~END
\endsourcecode

The code described here is the skeleton of a program performing
algebraic manipulation on polynomial.  Such a program must include a
call to {\tt BINOM} in order to initialize the table of binomial
coefficients.  

In order to store the coefficient of a monomial with
exponents {\tt J} (an integer array with dimension {\tt NLIB} the user
must include a statement like

\sourcecode
~~~~~~~K = INDICE(J,NLIB)
\endsourcecode

\noindent
and then say, e.g., {\tt F(K)=$\ldots$} which stores the coefficient
at the address {\tt K} of the array {\tt F}.

Suppose instead that we must perform an operation on all coefficients
of degree {\tt IORD} of a given function {\tt F}.  We need to perform
a loop on all the corresponding indices and retrieve the corresponding
exponents.  Here is a sketch of the code.

\sourcecode
C~~~~~~Compute the minimum and maximum index NMIN and NMAX
C~~~~~~of the coefficients of order IORD.
C
~~~~~~~IB=IORD+NMEN1
~~~~~~~NMIN = IBIN(IB,NLIB)
~~~~~~~IB=IORD+NLIB
~~~~~~~NMAX = IBIN(IB,NLIB) - 1
C
C~~~~~~Loop on all coefficients
C
~~~~~~~DO 1 N = NMIN,NMAX
~~~~~~~CALL ESPON(N,J,NLIB)
~~~~~~~...~{\rm more code to operate on the coefficient}~F(N)~...
~1~~~~~CONTINUE
 \endsourcecode

\noindent
Let us add a few words of explanation.  According
to~\frmref{indexf.100}, the index of the first coefficient of degree
$s$ in $n$ variables is $I(s,0,\ldots,0)=N(n,s-1)$, and we also have
$N(n,s-1)=\left({n+s-1}\atop{n}\right)$ as explained at the end of
sect.~\sbsref{indexf.1.3}.  This explains how the limits {\tt NMIN}
and {\tt NMAX} are calculated as $N(n,s-1)$ and $N(n,s+1)-1$,
respectively.  The rest of the code is the loop that retrieves the
exponents corresponding to the coefficient of index {\tt N}.  


\section{indexf.2}{Trigonometric polynomials}
Let us now consider the more general case $\Ascr_n=\interi^n$.  The
index $n$ in $\Ascr_n$ denotes again the dimension of the space.  The
name used in the title of the section is justified because this case
occurs precisely in the representation of trigonometric polynomials,
as explained in sect.~\sbsref{funrep.1.2}.
 
We shall now denote by $|k|=|k_1|+\ldots+|k_n|$ the length (or norm)
of the vector $k\in\interi^n$.  The tail $t(k)$ of a vector $k$ will be
defined again as $t(k)=(k_2,\ldots,k_n)$.

\subsection{indexf.2.1}{Ordering relation}
Pick a fixed $n$, and consider the finite
family of sets $\Ascr_{1}=\interi,\ldots,\Ascr_{n}=\interi^n$.

\blankq
\begingroup
\noindent\enufnt
Let $k,k'\in\Ascr_m$, with any $1\le m\le n$.  We shall say $k'\prec k$
in case one of the following conditions is true:

\item{(i)}$m\ge 1\ \wedge\ |k'|\lt |k|\>${\rm ;}
\item{(ii)}$m\gt 1\ \wedge\ |k'|=|k|\ \wedge\ |k'_1|\gt |k_1|\>${\rm ;}
\item{(iii)}$m\ge 1\ \wedge\ |k'|=|k|\ \wedge\ |k'_1|=|k_1|
                        \ \wedge\ k'_1\gt k_1\>${\rm ;}
\item{(iv)}$m\gt 1   \ \wedge\ |k'|=|k|\ \wedge\ k'_1=k_1
                          \ \wedge\ t(k')\prec t(k)\>$.  

\endgroup

\table{indexf.2}{Ordering of integer vectors in $\interi^m$ 
for $m=2,3,4$.}  
{ 
   \begingroup
\def\phm{{\phantom{-}}}
$$
\vcenter{\openup1\jot\halign{
\quad\hfil$\displaystyle{#}$\quad
&\quad\hfil$\displaystyle{#}$\hfil\quad
&\quad\hfil$\displaystyle{#}$\hfil\quad
&\quad\hfil$\displaystyle{#}$\hfil\quad
\cr
I(k)    &  m=2            &  m=3                   &  m=4                          \cr
\noalign{\vskip 2pt}
\noalign{\hrule}
\noalign{\vskip 3pt}
0\quad  & (\phm 0,\phm 0) & (\phm 0,\phm 0,\phm 0) & (\phm 0,\phm 0,\phm 0,\phm 0) \cr
1\quad  & (\phm 1,\phm 0) & (\phm 1,\phm 0,\phm 0) & (\phm 1,\phm 0,\phm 0,\phm 0) \cr
2\quad  & (    -1,\phm 0) & (    -1,\phm 0,\phm 0) & (    -1,\phm 0,\phm 0,\phm 0) \cr
3\quad  & (\phm 0,\phm 1) & (\phm 0,\phm 1,\phm 0) & (\phm 0,\phm 1,\phm 0,\phm 0) \cr
4\quad  & (\phm 0,    -1) & (\phm 0,    -1,\phm 0) & (\phm 0,    -1,\phm 0,\phm 0) \cr
5\quad  & (\phm 2,\phm 0) & (\phm 0,\phm 0,\phm 1) & (\phm 0,\phm 0,\phm 1,\phm 0) \cr
6\quad  & (    -2,\phm 0) & (\phm 0,\phm 0,    -1) & (\phm 0,\phm 0,    -1,\phm 0) \cr
7\quad  & (\phm 1,\phm 1) & (\phm 2,\phm 0,\phm 0) & (\phm 0,\phm 0,\phm 0,\phm 1) \cr
8\quad  & (\phm 1,    -1) & (    -2,\phm 0,\phm 0) & (\phm 0,\phm 0,\phm 0,    -1) \cr
9\quad  & (    -1,\phm 1) & (\phm 1,\phm 1,\phm 0) & (\phm 2,\phm 0,\phm 0,\phm 0) \cr
10\quad & (    -1,    -1) & (\phm 1,    -1,\phm 0) & (    -2,\phm 0,\phm 0,\phm 0) \cr
11\quad & (\phm 0,\phm 2) & (\phm 1,\phm 0,\phm 1) & (\phm 1,\phm 1,\phm 0,\phm 0) \cr
12\quad & (\phm 0,    -2) & (\phm 1,\phm 0,    -1) & (\phm 1,    -1,\phm 0,\phm 0) \cr
13\quad & (\phm 3,\phm 0) & (    -1,\phm 1,\phm 0) & (\phm 1,\phm 0,\phm 1,\phm 0) \cr
14\quad & (    -3,\phm 0) & (    -1,    -1,\phm 0) & (\phm 1,\phm 0,    -1,\phm 0) \cr
15\quad & (\phm 2,\phm 1) & (    -1,\phm 0,\phm 1) & (\phm 1,\phm 0,\phm 0,\phm 1) \cr
16\quad & (\phm 2,    -1) & (    -1,\phm 0,    -1) & (\phm 1,\phm 0,\phm 0,    -1) \cr
17\quad & (    -2,\phm 1) & (\phm 0,\phm 2,\phm 0) & (    -1,\phm 1,\phm 0,\phm 0) \cr
18\quad & (    -2,    -1) & (\phm 0,    -2,\phm 0) & (    -1,    -1,\phm 0,\phm 0) \cr
19\quad & (\phm 1,\phm 2) & (\phm 0,\phm 1,\phm 1) & (    -1,\phm 0,\phm 1,\phm 0) \cr
20\quad & (\phm 1,    -2) & (\phm 0,\phm 1,    -1) & (    -1,\phm 0,    -1,\phm 0) \cr
21\quad & (    -1,\phm 2) & (\phm 0,    -1,\phm 1) & (    -1,\phm 0,\phm 0,\phm 1) \cr
22\quad & (    -1,    -2) & (\phm 0,    -1,    -1) & (    -1,\phm 0,\phm 0,    -1) \cr
23\quad & (\phm 0,\phm 3) & (\phm 0,\phm 0,\phm 2) & (\phm 0,\phm 2,\phm 0,\phm 0) \cr
24\quad & (\phm 0,    -3) & (\phm 0,\phm 0,    -2) & (\phm 0,    -2,\phm 0,\phm 0) \cr
\cdots\quad&\cdots        & \cdots                 & \cdots                        \cr
\noalign{\vskip 2pt}
\noalign{\hrule}
}}
$$
\endgroup
  
}

\blankq\noindent
In table~\tabref{indexf.2} the order resulting from this definition
is illustrated for the cases $m=2,3,4$.

\blankq
If $m=1$ this ordering coincides with the ordering in $\interi$
introduced in sect~\secref{3}.  For $m\gt 1$, if (i),
(ii) and (iii) do not apply, then (iv) means that one must decrease the
dimension $n$ by replacing $k$ with its tail $t(k)$, and retry the
comparison.  Eventually, one ends up with $m=1$, falling back to the
one dimensional case to which only (i) and (iii) apply.

The ordering in this section has been defined for the case
$\Ascr_n=\interi^n$.  However, it will be useful to consider
particular subsets of $\interi^n$.  The natural choice will be to use
again the ordering relation defined here.  For example, the case of
integer vectors with non negative components discussed in
sect.~\sbsref{indexf.1.1} can be considered as a particular case: the
restriction of the ordering relation to that case gives exactly the
order introduced in sect.~\sbsref{indexf.1.1}.  Just remark that the
condition (iii) above becomes meaningless in that case, so that it can
be removed.

The set $\Pscr_n(k)$ of the elements preceding $k\in\Ascr^n$ in the
order above is defined as in sect.~\sbsref{indexf.1.1}.  Following the
line of the discussion in that section it is also convenient to give
some more definitions.  Pick a vector $k\in\Ascr_n$, and define the
sets $\Bscr_n^{(i)}(k)$, $\Bscr_n^{(ii)}(k)$, $\Bscr_n^{(iii)}(k)$ and
$\Bscr_n^{(iv)}(k)$ as the subsets of $\Ascr_n$ satisfying (i), (ii),
(iii) and (iv), respectively, in the ordering algorithm above.
Formally,
$$
\eqalign{
\Bscr_n^{(i)}(0)
&=\Bscr_n^{(ii)}(0)=\Bscr_n^{(iii)}(0)=\Bscr_n^{(iv)}(0)
 =\Bscr_1^{(ii)}(k)=\Bscr_1^{(iv)}(k)=\emptyset\ ,
\cr
\Bscr_n^{(i)}(k)
&=\{k'\in\Ascr_n\>:\>|k'|\lt |k|\}\ ,
\cr
\Bscr_n^{(ii)}(k)
&=\{k'\in\Ascr_n\>:\>|k'|=|k|\ \wedge\ |k'_1|\gt |k_1|\}\ ,
\cr
\Bscr_n^{(iii)}(k)
&=\{k'\in\Ascr_n\>:\>|k'|=|k|\ \wedge\ |k'_1|=|k_1|
\ \wedge\ k'_1\gt k_1\}\ ,
\cr
\Bscr_n^{(iv)}(k)
&=\{k'\in\Ascr_n\>:\>|k'|=|k|\ \wedge\ k'_1=k_1
\ \wedge\ t(k')\lt t(k)\}\ .
\cr
}
\formula{indexf.17}
$$
The sets $\Bscr_n^{(i)}(k)$, $\Bscr_n^{(ii)}(k)$, $\Bscr_n^{(iii)}(k)$
and $\Bscr_n^{(iv)}(k)$ are pairwise disjoint, and moreover
$$
\Bscr_n^{(i)}(k)\cup \Bscr_n^{(ii)}(k) \cup \Bscr_n^{(iii)}(k) 
\cup \Bscr_n^{(iv)}(k)
=\Pscr(k)\ .
$$
This easily follows from the definition.

\subsection{indexf.2.2}{Indexing function for trigonometric polynomials}
Let $k\in\Ascr_n$.  In view of the definitions and of the properties
above the index function, defined as in~\frmref{indexf.0}, turns out
to be
$$
I(0)=0\ ,\quad
I(k)= \#\Bscr_n^{(i)}(k) +\#\Bscr_n^{(ii)}(k) 
      +\#\Bscr_n^{(iii)}(k) +\#\Bscr_n^{(iv)}(k)\ .
\formula{indexf.108}
$$

Let us introduce the $J$-table and the $N$-table as
$$
\eqalign{
J(n,s)&=\#\{k\in\Ascr_n\>:\>|k|=s\}\ ,\cr
N(n,s)&=\sum_{j=0}^{s}J(n,j)\qquad\qquad\qquad{\rm for\ }n\ge
1\,,\>s\ge 0\ .\cr
}
\formula{indexf.110}
$$
We claim that the index function can be recursively computed as
$$
\eqalign{
I(0)&=0\ ,
\cr
I(k)&=\left\{
\vcenter{\openup1\jot\halign{
$\displaystyle{#}$\hfil
&\quad{\rm for\ }$\displaystyle{#}$\hfil
\cr
N(n,|k|-1)  & |k_1|=|k|\ \wedge\ k_1\ge 0\ ,
\cr
N(n,|k|-1)+1 & |k_1|=|k|\ \wedge\ k_1\lt 0\ ,
\cr
N(n,|k|-1)+N(n-1,|k|-|k_1|-1)+I(t(k)) 
   & |k_1|\lt |k|\ \wedge\ k_1\ge 0\ ,
\cr
N(n,|k|-1)+N(n-1,|k|-|k_1|)+I(t(k)) 
   & |k_1|\lt |k|\ \wedge\ k_1\lt 0\ .
\cr
}}\right.
\cr
}
\formula{indexf.200}
$$
This formula follows from
$$
\formdef{indexf.120}
\formdef{indexf.121}
\formdef{indexf.122}
\formdef{indexf.123}
\leqalignno{
\qquad\#\Bscr_n^{(i)}(k)
&=N(n,|k|-1)\ ;
&\frmref{indexf.120}
\cr
\qquad\#\Bscr_n^{(ii)}(k)
&=\left\{\vcenter{\openup1\jot\halign{
  \hbox to 12pc{$\displaystyle{#}$\hfil}\hfil
   &\quad{\rm for\ }$\displaystyle{#}$\hfil
     \cr
  0 & |k_1|=|k|\ ,
   \cr
  2 N(n-1,|k|-|k_1|-1) & |k_1|\lt |k|\ ;
   \cr
  }}\right.
&\frmref{indexf.121}
\cr
\qquad\#\Bscr_n^{(iii)}(k)
&=\left\{\vcenter{\openup1\jot\halign{
  \hbox to 12pc{$\displaystyle{#}$\hfil}\hfil
  &\quad{\rm for\ }$\displaystyle{#}$\hfil
  \cr
  0 & |k_1|\le |k|\ \wedge\ k_1\ge 0\ ,
   \cr
  J(n-1,|k|-|k_1|) & |k_1|\le |k|\ \wedge\ k_1\lt 0\ ;
   \cr
  }}\right.
&\frmref{indexf.122}
\cr
\qquad\#\Bscr_n^{(iv)}(k)
&=\left\{\vcenter{\openup1\jot\halign{
  \hbox to 12pc{$\displaystyle{#}$\hfil}\hfil
  &\quad{\rm for\ }$\displaystyle{#}$\hfil
  \cr
  I(t(k)) & |k_1|=|k|\ ,
   \cr
  I(t(k))-N(n-1,|k|-|k_1|-1) & |k_1|\lt|k|\ .
   \cr
}}\right.
&\frmref{indexf.123}
\cr
}
$$
The equality~\frmref{indexf.120} is a straightforward consequence of
the definition~\frmref{indexf.17}.  The equality~\frmref{indexf.121}
follows by remarking that for $|k_1|=|k|$ we have
$\Bscr_n^{(ii)}(k)=\emptyset$, and for $|k_1|\lt |k|$ we have
$$
\Bscr_n^{(ii)}(k)=B_n^{+}(k)\cup B_n^{-}(k)\ ,\quad
B_n^{+}(k)\cap B_n^{-}(k)=\emptyset\ ,
$$
with
$$
\eqalign{
B_n^{+}(k)
&=\bigcup_{0\le l\lt |k|-|k_1|}
   \{k'\in\Ascr_n\>:\>k'_1=|k|-l\ \wedge\ |t(k)|=l\}\ ,
\cr
B_n^{-}(k)
&=\bigcup_{0\le l\lt |k|-|k_1|}
   \{k'\in\Ascr_n\>:\>k'_1=l-|k|\ \wedge\ |t(k)|=l\}\ ;
\cr
}
$$
use also $\#B_n^{+}(k)=\#B_n^{-}(k)$.
The equality~\frmref{indexf.122} follows from
$$
\Bscr_n^{(iii)}(k)=\left\{
\vcenter{\openup1\jot\halign{
$\displaystyle{#}$\hfil
&\quad{\rm for\ }$\displaystyle{#}$\hfil
\cr
\emptyset & |k_1|=|k|\ \wedge\ k_1\ge 0\ ,
\cr
\{k'\in\Ascr_n\>:\>k'_1=|k_1|\ \wedge\ |t(k')|=|k|-|k_1|\}
&|k_1|=|k|\ \wedge\ k_1\lt 0\ .
\cr
}}\right.
$$
Coming to~\frmref{indexf.123}, remark that
$$
\Bscr_n^{(iv)}(k)=
\{k'\in\Ascr_n\>:\> |t(k')|=|k|-|k_1|\ \wedge\ t(k')\prec t(k)\}\ .
$$
Proceeding as in the polynomial case we find again 
$$
\#\Bscr_n^{(iv)}(k)=\#\{\lambda\in\Ascr_{n-1}\>:\>|\lambda|=|k|-|k_1|
\ \wedge\ \lambda\prec t(k)\}\ ,
$$
and~\frmref{indexf.123} follows by remarking that 
$$
\Pscr_{n-1}(t(k))=
 \{\lambda\in\Ascr_{n-1}\>:\>|\lambda|=|k|-|k_1|
  \ \wedge\ \lambda\prec t(k)\}
   \cup
    \{\lambda\in\Ascr_{n-1}\>:\>|\lambda|\lt |k|-|k_1|\}\ .
$$
Adding up all contributions~\frmref{indexf.200} follows.

\subsection{indexf.2.3}{Construction of the tables}
We show now how to construct recursively the $J$-table, so that the
$N$-table can be constructed, too.   For $n=1$ we have, trivially,
$J(1,0)=1$ and $J(1,s)=2$ for $s\gt 0$.   For $n\gt 1$ use the
elementary property
$$
\{k\in\Ascr_n\>:\>|k|=s\}= 
 \bigcup_{-s\le j\le s}\{k\in\Ascr_n\>:\>k_1=j
  \ \wedge\ |t(k)|=s-|j|\}\ .
$$
Therefore 
$$
\eqalign{
J(1,0)&=1\ ,
\cr
J(1,s)&=2\ ,
\cr
J(n,s)&=\sum_{j=-s}^{s}J(n-1,s-|j|)\quad {\rm for\ }n\gt 1\ .
\cr
}
\formula{indexf.211}
$$
This completely determines the $J$-table.

\subsection{indexf.2.4}{Inversion of the index function}
The problem is to find the vector $k$ of given dimension $n$
corresponding to the given index $l$.  For $n=1$ the function $I(k)$
and its inverse $I^{-1}(l)$ are given by~\frmref{indexf.0.1}
and~\frmref{indexf.0.2}.  Therefore in the rest of this section we shall
assume $n\gt 1$.  We shall give a recursive algorithm, showing how to
determine $k_1$ and $I\bigl(t(k)\bigr)$.

\item{(i)}If $l=0$ then $k=0$, and there is nothing else to do.

\item{(ii)}Assuming that $l\gt 0$, determine $s$ such that
$$
N(n,s-1)\le l \lt N(n,s)\ .
$$
\item{}From this we know that $|k|=s$.

\item{(iii)}Define $l_1=l-N(n,s-1)$, so that $I\bigl(t(k)\bigr)\le
l_1$ by~\frmref{indexf.200}.  If $l_1=0$ set $s_1=0$; else, determine
$s'$ such that
$$
N(n-1,s'-1)\le l_1 \lt N(n-1,s')\ ,
$$
and let $s_1=\min(s',s)$.  In view of $I\bigl(t(k)\bigr)\le l_1$ we
know that $|t(k)|\le s_1$.  Remark also that $s_1=0$ if and only if
$l_1=0$.  For, if $s_1\ge 1$ then we have $l_1\ge N(n-1,0)=1$.

\item{(iv)}If $l_1=0$, then by the first of~\frmref{indexf.200} we
conclude 
$$
k_1=|k|=s\ ,\quad t(k)=0\ ,
$$
and there is nothing else to do.

\item{(v)}If $l_1=1$, then by the second of~\frmref{indexf.200} we
conclude 
$$
k_1= -|k|= -s\ ,\quad t(k)=0\ ,
$$
\item{}and there is nothing else to do.

\item{(vi)}If $l_1\gt 1$ and $s_1\gt 0$, we first look if we 
can set $0\le k_1\lt |k|$.  In view of the third of~\frmref{indexf.200}
we should have
$$
|k|-k_1=s_1\ ,\quad
\bigl|t(k)\bigr|=s_1\ ,\quad
I\bigl(t(k)\bigr)=l_1-N(n-1,s_1-1)\ .
$$
\item{}This can be consistently made provided the conditions
$$
s_1\gt 0\quad {\rm and}\quad I\bigl(t(k)\bigr)\ge N(n-1,s_1-1)
$$
\item{}are fulfilled.  The condition $s\gt 0$ is already satisfied.
By~\frmref{indexf.200}, the second condition is fulfilled provided
$l_1\ge 2 N(n-1,s_1-1)$.  This has to be checked.  
\itemitem{(vi.a)}If the second condition is true, then set
$k_1=|k|-s_1$, and recall that $\bigl|t(k)\bigr|=s_1$.  Hence, we can
replace $n$, $l$, and $s$ by $n-1$, $l_1-N(n-1,s_1-1)$ and $s_1$,
respectively, and proceed by recursion restarting again from the point
(iii).
\itemitem{(vi.b)}If the second condition is false, then we proceed with
the next point.

\item{(vii)}Recall that $l_1\gt 1$, and remark that we have also
$s_1\gt 1$.  Indeed, we already know $s_1\gt 0$, so we have to exclude
the case $s_1=1$.  Let, by contradiction, $s_1=1$.   Then we have
$l_1\ge 2= 2N(n-1,s_1-1)$, which is the case already excluded
by~(vi).  We conclude $s_1\gt 1$.  We look now for the possibility of
setting $|k_1|\lt |k|$ and $k_1\lt 0$.  In view of the fourth
of~\frmref{indexf.200} we should have
$$
|k|+k_1=s_1-1\ ,\quad
\bigl|t(k)\bigr|=s_1-1\ ,\quad
I\bigl(t(k)\bigr)=l_1-N(n-1,s_1-1)\ .
$$
\item{}This can be consistently made provided the conditions
$$
s_1\gt 1\quad {\rm and}\quad I\bigl(t(k)\bigr)\ge N(n-1,s_1-2)
$$
\item{}are fulfilled.  The condition $s_1\gt 1$ is already satisfied.  
As to the second condition, by~\frmref{indexf.200} it is fulfilled
provided $l_1\gt N(n-1,s_1-1)+N(n-1,s_1-2)$.  This has to be checked.
\itemitem{(vii.a)}If the second condition is true, then set
$k_1=-|k|+s_1-1$, and recall that $\bigl|t(k)\bigr|=s_1-1$.  
Hence, we can
replace $n$, $l$, and $s$ by $n-1$, $l_1-N(n-1,s_1-2)$ and $s_1-1$,
respectively, and proceed by recursion restarting again from the point
(iii).
\itemitem{(vii.b)}If the second condition is false we must decrease 
$s_1$ by one and start again with the point (vi); remark that $s_1\gt
1$ implies $s_1-1\gt 0$, which is the first of the two conditions to
be satisfied at the point~(vi), hence the recursion is correct.  

\noindent
Since $l_1\gt 1$ we have $l_1\gt 2N(n-1,0)$, so that the conditions of
point~(vi) are satisfied for $s=1$.  Hence the algorithm above does not
fall into an infinite loop between points~(vi) and~(vii).  On the other
hand, for $n=1$ either~(iii) or~(iv) applies, so that the algorithm
stops at that point.

\section{4}{Storing the coefficients for sparse functions}
The method of storing the coefficient using the index, as illustrated
in sect.~\secref{2}, is the most direct one, but reveals to be
ineffective when most of the coefficients of a function are zero
(sparse function).  For, allocating memory space for all coefficients
results in a wasting of memory.

A method that we often use is to store the coefficients using a tree
structure based on the index.  However we should warn the reader that the method described
here has the advantage of being easily programmed, but does not
pretend to be the most effective one.   Efficient programming of tree
structure is described, e.g., in the monumental books {\corsivo The
art of computing programming}, by D.E.~Knuth~\dbiref{Knuth-1968}.

\subsection{4.1}{The tree structure}
The first information we need is how many bits are needed in order to
represent the maximum index for a function.  We shall refer to this
number as the {\corsivo length of the index}.  In the scheme that we
are presenting here this is actually the length of the path from the
root of the tree to its leave, where the coefficient is found.

\figure{laplat.1}{\psfig{figure=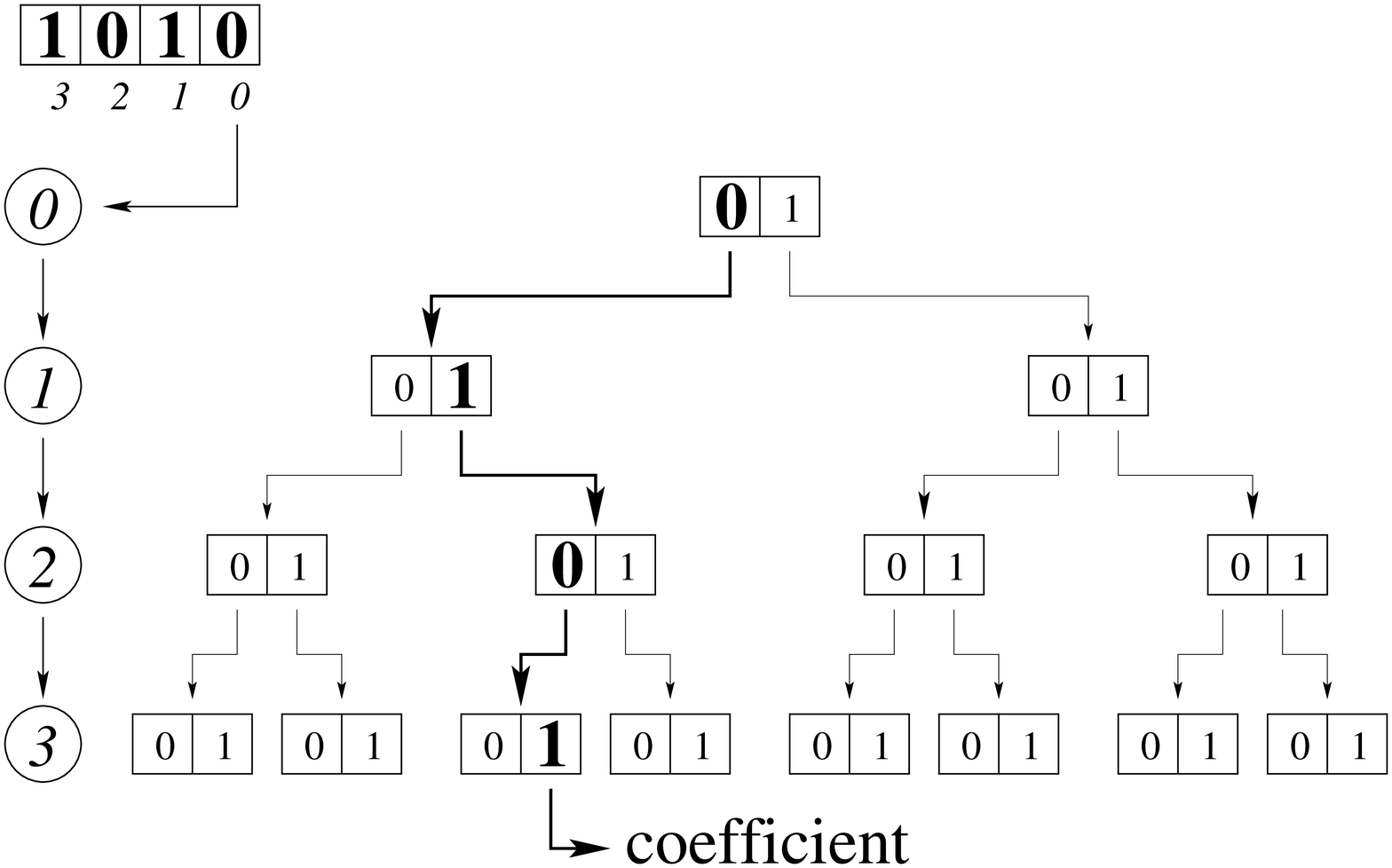,width=14 truecm}}{Illustrating
the tree structure for a 4-bit long index (see text).}

In fig.~\figref{laplat.1} we illustrate the scheme assuming that 4
bits are enough, i.e., there are at most 16 coefficients indexed from
$0$ to $15$.  The case is elementary, of course, but the method is the
general one, and is extended to, e.g., several millions of
coefficients (with a length a little more than 20) in a
straightforward manner.  The bits are labeled by their position,
starting from the less significant one (choosing the most significant
one as the first bit is not forbidden, of course, and sometimes may be
convenient).  The label of the bit corresponds to a level in the tree
structure, level $0$ being the root and level $3$ being the last one,
in our case.  At level zero we find a cell containing two pointers,
corresponding to the digit $0$ and $1$, respectively.  To each digit
we associate a cell of level $1$, which contains a pair of pointers,
and so on until we reach the last level ($3$ in our case).  Every number that
may be represented with $4$ bits generates a unique path along the
tree, and the last cell contains pointers to the coefficient.  The
example in the figure represents the path associated with the binary
index $1010$, namely $10$ in decimal notation.

\figure{laplat.2}{\psfig{figure=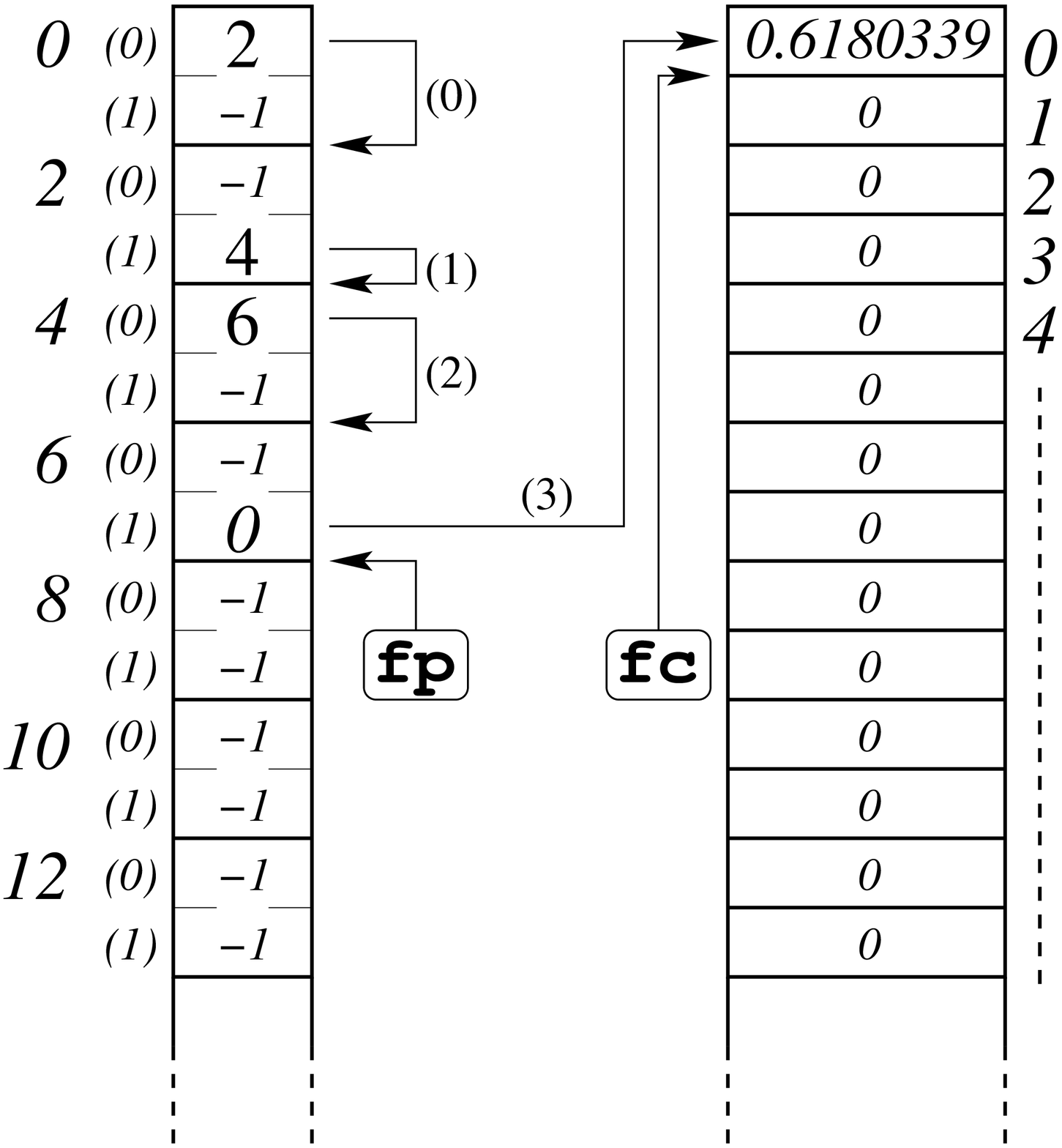,height=12 truecm}}{Illustrating
how the tree structure is stored in memory (see text).}

Let us also illustrate how this structure may be represented in
memory, trying to avoid wasting of space.  We use two separate arrays,
the first one for pointers and the second one for the coefficients, as
illustrated in fig.~\figref{laplat.2}.  The cells containing pairs of
pointers are allocated in the first array, the root of the tree having
label zero.  The label of a cell is always even: the first element
corresponds to the zero bit, the next one (with odd label) to the bit
one.

The arrays are initially allocated with appropriate size, and are
cleared.  A good method is to fill the array of pointers with $-1$
(denoting an unused pointer) and the coefficients table with zeros.
We also keep track of the first unused cell in the array, which
initially is set to $2$ because the root cell is considered to be in
use, and of the first free coefficient, which initially is $0$.

We shall use the following notations: ${\tt cell}(2j)$ is the cell with
even label $2j$ in the array; ${\tt cell}(2j,0)$ and ${\tt cell}(2j,1)$
are the pointers corresponding to a bit 0 or 1 which are stored at
locations $2j$ and $2j+1$, respectively, in the array of pointers; ${\tt
coef}(j)$ is the $j$-th element of the array of coefficients; ${\tt
cc}$ is the current cell and ${\tt cb}$ is the current bit (see below
for the meaning); ${\tt fp}$ is the label of the first free (unused)
cell of pointers; ${\tt fc}$ is the label of the first free
coefficient; $\ell$ is the length of the index.

\subsection{4.2}{Storing the first coefficient}
Let us describe how the first coefficient is stored. Suppose we want
to store the value $x$ as the coefficient corresponding to a given
index.  Here is the scheme.

\item{(i)}{\corsivo Initialization:} set ${\tt cc}=0$ and ${\tt
cb}=0$.  The values of ${\tt fp}=2$ and ${\tt fc}=0$ have already been
set when during the array allocation.

\item{(ii)}{\corsivo Creating a path:} repeat the following steps
until ${\tt cb}$ equals $\ell-1$:

\itemitem{(ii.a)}if the bit at position ${\tt cb}$
in the index is 0, then redefine  ${\tt cell}({\tt
cc},0)={\tt fp}$; else redefine ${\tt cell}({\tt cc},1)={\tt fp}$;

\itemitem{(ii.b)}set ${\tt cc}={\tt fp}$ and increment ${\tt fp}$ by
$2$ (point to the next free cell);

\itemitem{(ii.c)}increment ${\tt cb}$ by $1$  (next bit).

\item{(iii)}{\corsivo Store the coefficient:} 

\itemitem{(iii.a)}if the bit at position ${\tt cb}$
in the index is 0, then redefine  ${\tt cell}({\tt
cc},0)={\tt fc}$; else redefine ${\tt cell}({\tt cc},1)={\tt fc}$;

\itemitem{(iii.b)}set ${\tt coef}({\tt fc})=x$;

\itemitem{(iii.c)}increment ${\tt fc}$ by $1$ (point to the next free
coefficient).

\noindent
Programming this algorithm in a language such as {\tt C} or {\tt
FORTRAN} requires some $10$ to $20$ statements.

Let us see in detail what happens if we want to store the coefficient
$0.6180339$ with index $1010$ and $\ell=4$, as illustrated in
fig.~\figref{laplat.2}.  Here is the sequence of operations actually
made
$$
\vcenter{\openup1\jot\halign{
{#}\hfil
&\quad$\displaystyle{#}$\ ,\hfil
&\quad$\displaystyle{#}$\ ,\hfil
&\quad$\displaystyle{#}$\ ,\hfil
&\quad$\displaystyle{#}$\hfil
&\quad {#}\hfil
\cr
step\ (i): & {\tt cc}=0 & {\tt cb}=0 & {\tt fp}=2 & {\tt fc}=0\ ;
\cr
step\ (ii): & {\tt cell(0,0)}=2   & {\tt cc}=2  & {\tt fp}=4  & {\tt cb}=1\ ,& then\ ,
\cr
            & {\tt cell(2,1)}=4   & {\tt cc}=4  & {\tt fp}=6  & {\tt cb}=2\ ,& then\ ,
\cr
            & {\tt cell(4,0)}=6   & {\tt cc}=6  & {\tt fp}=8  & {\tt cb}=3\ ,& end of loop\ ;
\cr
step\ (iii): & {\tt cell(6,1)}=0  & \multispan 2\quad{\tt coef}(0)=0.6180339\ , & {\tt fc}=1\ ,  & end of game\ .
\cr
}}
$$
After this, the contents of the arrays are as represented in fig.~\figref{laplat.2}.

\subsection{4.2}{Retrieving a coefficient}
The second main operation is to retrieve a coefficient, which possibly has never been
stored.  In the latter case, we assume that the wanted coefficient is zero.  Here is a
scheme.

\item{(i)}{\corsivo Initialization:} set ${\tt cc}=0$ and ${\tt cb}=0$.

\item{(ii)}{\corsivo Follow a path:} repeat the following steps
until ${\tt cb}$ equals $\ell$:

\itemitem{(ii.a)}save the current value of ${\tt cc}$;

\itemitem{(ii.b)}if the bit at position ${\tt cb}$
in the index is 0, then redefine ${\tt cc}$ as  ${\tt cell}({\tt
cc},0)$; else redefine ${\tt cc}$ as  ${\tt cell}({\tt cc},1)$;

\itemitem{(ii.c)}if ${\tt cc}=-1$ then the coefficient is undefined.
Return $0$ as the value of the coefficient;

\itemitem{(ii.d)}increment ${\tt cb}$ by $1$  (next bit).

\item{(iii)}{\corsivo Coefficient found:} return 
the coefficient ${\tt coef}({\tt cc})$.

Let us give a couple of examples in order to better illustrate the
algorithm.  Suppose that we are looking for the coefficient
corresponding to the binary index $1010$.  By following the algorithm
step by step, and recalling that in our example the length of the
index is $4$, the reader should be able to check that the sequence of
operations is the following:
$$
\vcenter{\openup1\jot\halign{
{#}\quad\hfil
&$\displaystyle{#}$\hfil
&\quad$\displaystyle{#}$\hfil
&\quad {#}\hfil
\cr
step\ (i): & {\tt cc}=0\ , & {\tt cb}=0\ ;
\cr
step\ (ii): & {\tt cc}=2\ , & {\tt cb}=1\ ,& then\ ,
\cr
            & {\tt cc}=4\ , & {\tt cb}=2\ ,& then\ ,
\cr
            & {\tt cc}=6\ , & {\tt cb}=3\ ,& then\ ,
\cr
            & {\tt cc}=0    & {\tt cb}=4\ ,& end of path\ ;
\cr
step\ (iii): &  &                        & return\ $0.6180339$\ .
\cr
}}
$$
The returned value is that of ${\tt coef}(0)$, stored in the location $0$ of the
coefficients array.

Suppose now that we are looking for the coefficient corresponding to the binary index
$1110$.  Here is the actual sequence of operations:
$$
\vcenter{\openup1\jot\halign{
{#}\quad\hfil
&$\displaystyle{#}$\hfil
&\quad$\displaystyle{#}$\hfil
&\quad {#}\hfil
\cr
step\ (i): & {\tt cc}=0\ , & {\tt cb}=0\ ;
\cr
step\ (ii): & {\tt cc}=2\ , & {\tt cb}=1\ ,& then\ ,
\cr
            & {\tt cc}=4\ , & {\tt cb}=2\ ,& then\ ,
\cr
            & {\tt cc}=-1\ , & {\tt cb}=2\ ,& return zero\ .
\cr
}}
$$
Here the algorithm stops because a coefficient has not been found. 

\subsection{4.4}{Other operations}
Having implemented the two operations above, the reader should be able to implement also
the following operations:

\item{(i)}storing a new coefficient corresponding to a given index;

\item{(ii)}adding something to a given coefficient;

\item{(iii)}multiplying a given coefficient by a number.

\noindent
These are the basic operations that we need in order to perform an elementary computer
algebra.  Let us add a few hints.

\figure{laplat.3}{\psfig{figure=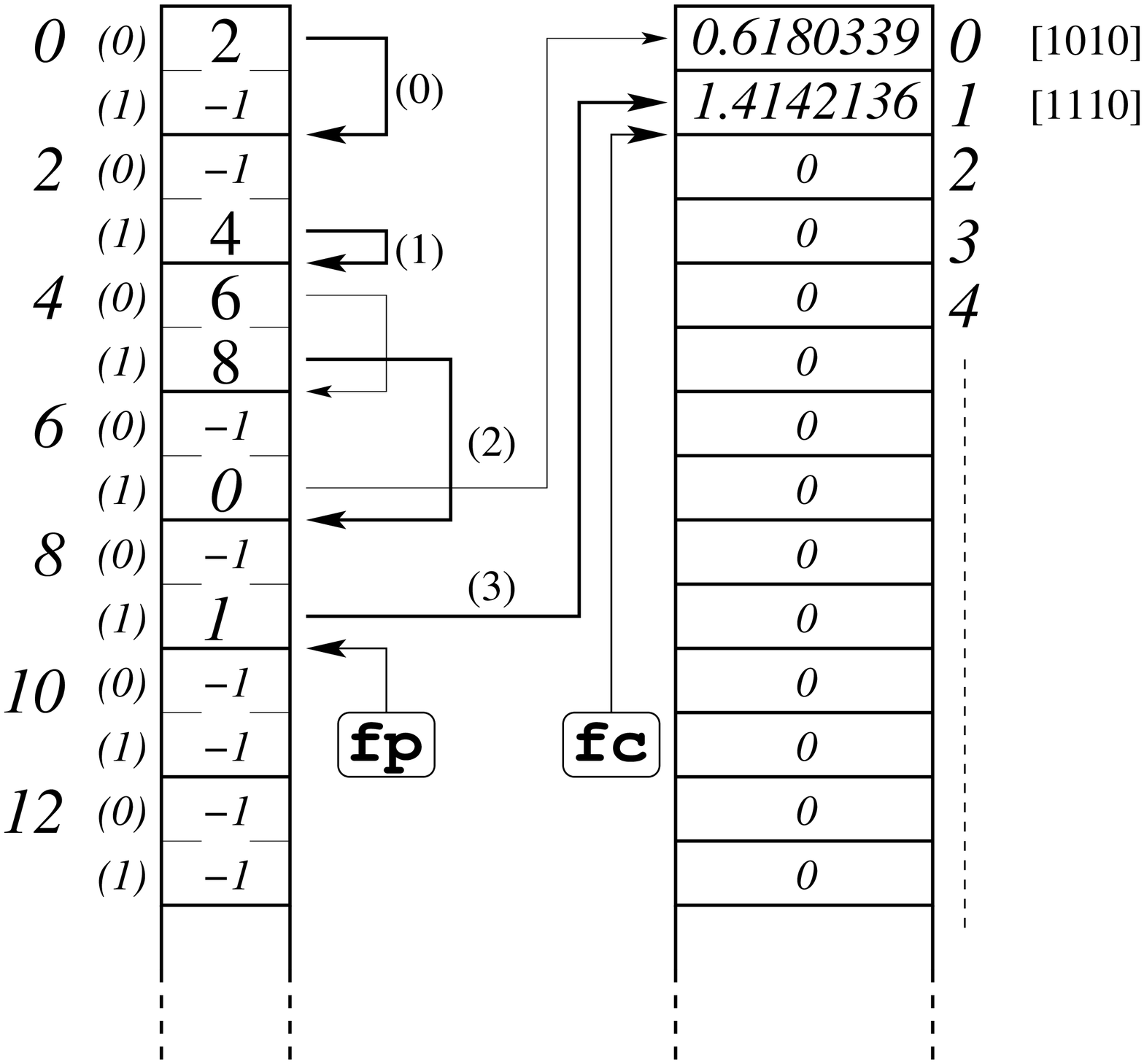,height=12 truecm}}{Inserting
a new coefficient in a tree structure (see text).}

Storing a new coefficient requires perhaps some moment of thinking.  Using the index,
one should follow the corresponding path in the tree (as in the operation of retrieving)
until either happens: the coefficient is found, or the search fails at some point.  If the
coefficient is found, then it can be overwritten if the new value has to replace the old
one.  On failure, the path must be completed by appropriately defining the pointers (as in
the case of the first coefficient), and then the coefficient can be stored in the
appropriate location.  As an exercise, suppose that we want to store the coefficient
$1.4142136$ corresponding to the binary index $1110$.  After completing the operation the
memory should look as in fig.~\figref{laplat.3}.

Adding something to a given coefficient is not very different from the previous
operation.  Just follow the path.  If the coefficient is found, then add the wanted value
to it.  On failure, just change the ``add'' operation to a ``store'' one, and proceed as in
the case~(i).

Multiplying a coefficient by a constant is even easier.  If the coefficient is
found, then do the multiplication.  On failure, just do nothing.

Further operations can be imagined, but we think that we have described the basic ones.
There are just a couple of remarks.

The method illustrated here uses an amount of memory that clearly depends on the number of
non zero coefficients of a function.  However, this amount is typically not known in
advance.  Thus, enough memory should be allocated at the beginning in order to assure that
there is enough room.  When a function is filled, and we know that it will not be changed,
the excess of memory can be freed and reused for other purposes.  Every operating system and
language provides functions that allow the programmer to allocate memory blocks and resize
them on need.

A second remark is that other storing methods can be imagined.  E.g., once a function is
entirely defined it may be more convenient to represent it as a sequential list of pairs
(index, coefficient).  This is definitely a very compact representation for a sparse
function (although not the best for a crowded one).

\section{5}{Applications}
We report here some examples of application of algebraic
manipulation that have been obtained by implementing the formal
algorithm of sect.~\secref{1.a}. We consider three cases, namely the
model of H\'enon and Heiles, the Lagrangian triangular equilibria for
the Sun-Jupiter system and the planetary problem including Sun,
Jupiter, Saturn and Uranus (SJSU).

\figure{hh.sez}{\vtop{\openup1\jot\halign{
\hfil{#}
&{#}\hfil
\cr
\noalign{\vskip 12pt}
\cr
$E=\frac{1}{100}$\hfil & \hfil $E=\frac{1}{12}$ \cr
{\psfig{figure=henf1.eps,height=7.5truecm}}
&{\psfig{figure=henf2.eps,height=7.5truecm}}
\cr
\noalign{\vskip 6pt}
$E=\frac{1}{8}$\hfil & \hfil $E=\frac{1}{6}$ \cr
{\psfig{figure=henf3.eps,height=7.5truecm}}
&{\psfig{figure=henf4.eps,height=7.5truecm}}
\cr
}}}{Poincar\'e sections for the H\'enon and Heiles model.  The energies
are as in the original paper.}

\figure{hh.1_100}{\vtop{\openup1\jot\halign{
\hfil{#}
&{#}\hfil
\cr
\noalign{\vskip 12pt}
\cr
{\psfig{figure=hen_sez4_en1_100.eps,height=7.5truecm}}
&{\psfig{figure=hen_sez8_en1_100.eps,height=7.5truecm}}
\cr
}}}{Level lines of the
first integral truncated at orders 4 and 8, for energy
$E={\scriptstyle\frac{1}{100}}$.  The figure for truncation orders up
to 58 are actually the same as for order 8.}

\figure{hh.1_12}{\vtop{\openup1\jot\halign{
\hfil{#}
&{#}\hfil
\cr
\noalign{\vskip 12pt}
\cr
{\psfig{figure=hen_sez8_en1_12.eps,height=7.5truecm}}
&{\psfig{figure=hen_sez32_en1_12.eps,height=7.5truecm}}
\cr
{\psfig{figure=hen_sez43_en1_12.eps,height=7.5truecm}}
&{\psfig{figure=hen_sez58_en1_12.eps,height=7.5truecm}}
\cr
}}}{Level lines of the
first integral truncated at orders 8, 32, 43 and 58, for energy
$E={\scriptstyle\frac{1}{12}}$.  A good correspondence with the
Poincar\'e sections is found at orders, roughly, 8 to 32.  Then the
level lines start to disprove, in agreement with the asymptotic
character of the series.}

\figure{hh.1_8}{\vtop{\openup1\jot\halign{
\hfil{#}
&{#}\hfil
\cr
\noalign{\vskip 12pt}
\cr
{\psfig{figure=hen_sez8_en1_8.eps,height=7.5truecm}}
&{\psfig{figure=hen_sez9_en1_8.eps,height=7.5truecm}}
\cr
{\psfig{figure=hen_sez10_en1_8.eps,height=7.5truecm}}
&{\psfig{figure=hen_sez27_en1_8.eps,height=7.5truecm}}
\cr
}}}{Level lines of the first integral truncated at orders 8, 9, 10 and
27, for energy $E={\scriptstyle\frac{1}{8}}$.  Some correspondence
with the Poincar\'e sections is found around the order 9.
Then the level lines are definitely worse, making even more evident
the asymptotic character of the series.}

\subsection{5.1}{The model of H\'enon and Heiles}
A wide class of canonical system with Hamiltonian of the form 
$$
H(x,y) = \frac{\omega_1}{2}(y_1^2+x_1^2) + 
          \frac{\omega_2}{2}(y_2^2+x_2^2)+x_1^2 x_2
\formula{espnum.5}
$$
has been studied by Contopoulos, starting at the end of the fifties,
for different values of the frequencies.  This approximates the motion
of a star in a galaxy, at different distances from the center.  A wide
discussion on the use of these models in galactic dynamics and on the
construction of the so called ``third integral'' can be found in the
book~\dbiref{Contopoulos-2002}.  The third integral is constructed as
a power series $\Phi=\Phi_2+\Phi_3+\ldots$ where $\Phi_s$ is a
homogeneous polynomial of degree $s$ which is the solution of the
equation $\left\{H,\Phi\right\}=0$, where $\{\cdot,\cdot\}$ is the
Poisson bracket (see, e.g.,~\dbiref{Whittaker-1916}
or~\dbiref{Contopoulos-1960}).  A different method is based on the
construction of the Birkhoff normal form~\dbiref{Birkhoff-1927}.

A particular case with two equal frequencies and Hamiltonian
$$
H(x,y) = \frac{1}{2}(y_1^2+x_1^2) + 
          \frac{1}{2}(y_2^2+x_2^2)+x_1^2 x_2 -\fraz{1}{3} x_2^3
\formula{hh.ham}
$$
has been studied by H\'enon and Heiles in 1964~\dbiref{Henon-1964}.
This work has become famous since for the first time the existence of
a chaotic behavior in a very simple system has been stressed, showing
some figures.  It should be remarked that the existence of chaos had
been discovered by Poincar\'e in his memory on the problem of three
bodies~\dbiref{Poincare-1889}, but it had been essentially forgotten.

A program for the construction of the third integral has been
implemented by Contopoulos since 1960.  He made several comparisons
between the level lines of the integral so found on the surface of
constant energy and the figures given by the Poincar\'e sections of
the orbits.  A similar calculation for the case of H\'enon and Heiles
has been made by Gustavson~\dbiref{Gustavson-1966}, who used the
normal form method.  The third integral was expanded up to order 8,
which may seem quite low today, but it was really difficult to do
better with the computers available at that time.  Here we reproduce
the figures of Gustavson extending the calculation up to order 58,
which is now easily reached even on a PC.

In fig.~\figref{hh.sez} we show the Poincar\'e sections for the
values of energy used by H\'enon and Heiles in their paper.  As
stressed by the authors, an essentially ordered motion is found for
$E\lt {\scriptstyle\frac{1}{12}}$, while the chaotic orbits become predominant at higher
energies.

The comparison with the level lines of the third integral at energy
$E={\scriptstyle\frac{1}{100}}$ is reported in
fig.~\figref{hh.1_100}.  The correspondence with the Poincar\'e
sections is evident even at order 8, as calculated also by Gustavson.
We do not produce the figures for higher orders because they are
actually identical with the one for order 8.  This may raise the hope
that the series for the first integral is a convergent one.

Actually, a theorem of Siegel states that for the Birkhoff normal form
divergence is a typical case~\dbiref{Siegel-1941}.  A detailed
numerical study has been made in~\dbiref{Contopolus-2003}
and~\dbiref{Contopolus-2004}, showing the mechanism of divergence.
Moreover, it was understood by Poincar\'e that perturbations series
typically have an asymptotic character (see~\dbiref{Poincare-1892},
Vol.~II).  Estimates of this type have been given, e.g.,
in~\dbiref{Giorgilli-1988.4} and~\dbiref{Giorgilli-1989}.

For energy $E={\scriptstyle\frac{1}{12}}$ (fig.~\figref{hh.1_12}) the asymptotic
character of the series starts to appear.  Indeed already at order 8
we have a good correspondence between the level lines and the
Poincar\'e section, as was shown also Gustavson's paper.  If we
increase the approximation we see that the correspondence remains good
up to order 32, but then the divergence of the series shows up, since
at order 43 an unwanted ``island'' appears on the right side of the
figure which has no correspondent in the actual orbits, and at order
58 a bizarre behavior shows up.

The phenomenon is much more evident for energy
$E={\scriptstyle\frac{1}{8}}$ (fig.~\figref{hh.1_8}).  Here some rough correspondence is
found around order 9, but then the bizarre behavior of the previous
case definitely appears already at order 27.

\figure{hh.norme}{\psfig{figure=hen_norme_gen.eps,width=13truecm}}{The
convergence radius evaluated with the ratio (left) and the root
(right) criterion.  In both cases the non convergence of the series is
evident.}

The non convergence of the normal form is illustrated in
fig.~\figref{hh.norme}.  Writing the homogeneous terms of degree $s$
of the third integral as $\Phi_s=\sum_{j,k}\phi_{j,k}x^jy^k$, we may
introduce the norm
$$
\bigl\|\Phi_{s}\bigr\| = \sum_{j,k} |\phi_{j,k}|\ .
$$
Then an indication of the convergence radius may be found by calculating
one of the quantities
$$
\bigl\|\Phi_{s}\bigr\|^{1/s}\ ,\quad
\frac{\bigl\|\Phi_{s}\bigr\|}{\bigl\|\Phi_{s-1}\bigr\|}\ ,\quad
\left(\frac{\bigl\|\Phi_{s}\bigr\|}{\bigl\|\Phi_{s-2}\bigr\|}\right)^{1/2}\ .
$$
The first quantity corresponds to the root criterion for power series.
The second one corresponds to the ratio criterion.  The third one is
similar to the ratio criterion, but in the present case turns out to
be more effective because it takes into account the peculiar behavior
of the series for odd and even degrees.  The values given by the root
criterion are plotted in the left panel of fig.~\figref{hh.norme}.
The data for the ratio criterion are plotted in the right panel, where
open dots and solid dots refer to the second and third quantities in the
formula above, respectively.  In all cases it is evident that the
values steadily increase, with no tendency to a definite limit.  The
almost linear increase is consistent with the behavior
$\bigl\|\Phi_{s}\bigr\|\sim s!$ predicted by the theory.

\subsection{5.2}{The Trojan asteroids}
The asymptotic behavior of the series lies at the basis of
Nekhoroshev theory on exponential stability.  The general result,
referring for simplicity to the case above, is that in a ball of
radius $\rho$ and center at the origin one has
$$
\bigl| \Phi(t)-\Phi(0)\bigr| \lt O(\rho^3) \quad
{\rm for}\ |t|\lt O(\exp(1/\rho^a))\ , 
$$
for some positive $a\le 1$.  This is indeed the result given by the
theory (see, e.g.,~\dbiref{Giorgilli-1988.4}).  In rough terms the idea
is the following.  Due to the estimate $\bigl\|\Phi_{s}\bigr\|\sim s!$
and remarking that $\dot\Phi=\bigl\{H,\Phi\bigr\}$ starts with terms
of degree $s+1$, one gets $\bigl|\dot\Phi\bigr| = O(s!\rho^{s+1})$.
Then one looks for an optimal degree $s$ which minimizes the time
derivative, i.e., $s\sim 1/\rho$.  By truncating the integrals at the
optimal order one finds the exponential estimate.

\figure{troiani}{\psfig{figure=tstab.eps,width=13truecm}}{The
estimated stability time and the optimal truncation order for the
$L_4$ point of the Sun-Jupiter system.}

However, the theoretical estimates usually give a value of $\rho$
which is useless in practical applications, being definitely too small.
Realistic results may be obtained instead if the construction of first
integrals for a given system if performed by computer algebra.  That
is, one constructs the expansion of the first integral up to an high
order, compatibly with the computer resources available, and then
looks for the optimal truncation order by numerical evaluation of the
norms.

The numerical optimization has been performed for the expansion of the
Hamiltonian in a neighborhood of the Lagrangian point $L_4$, in the
framework of the planar circular restricted problem of three bodies in
the Sun-Jupiter case.  This has a direct application to the dynamics
of the Trojan asteroids (see~\dbiref{Giorgilli-1997}).

The two first integrals which are perturbations of the harmonic actions
have been constructed up to order $34$ (close to the best possible
with the computers at that time).  The estimate of the time of
stability is reported in fig.~\figref{troiani}.  The lower panel gives
the optimal truncation order vs.\ $\log_{10}\rho$.   In the upper panel we
calculate the stability time as follows: for an initial datum inside a
ball of radius $\rho_0$ we determine the minimal time required for the
distance to increase up to $2\rho_0$.  Remark that the vertical scale
is logarithmic.  The units are chosen so that $\rho=1$ is the distance
of Jupiter from the Sun, and $t=2\pi$ is the period of Jupiter.
With this time unit the estimated age of the universe is about
$10^9$.  The figure shows that the obtained data are already
realistic, although, due to the unavoidable approximations, only four
of the asteroids close to $L_4$ known at the time of that work did
fall inside the region of stability for a time as long as the age of the
Universe.

\figure{sjsu}{\vtop{\openup1\jot\halign{
\hfil{#}\hfil
\cr
{\psfig{figure=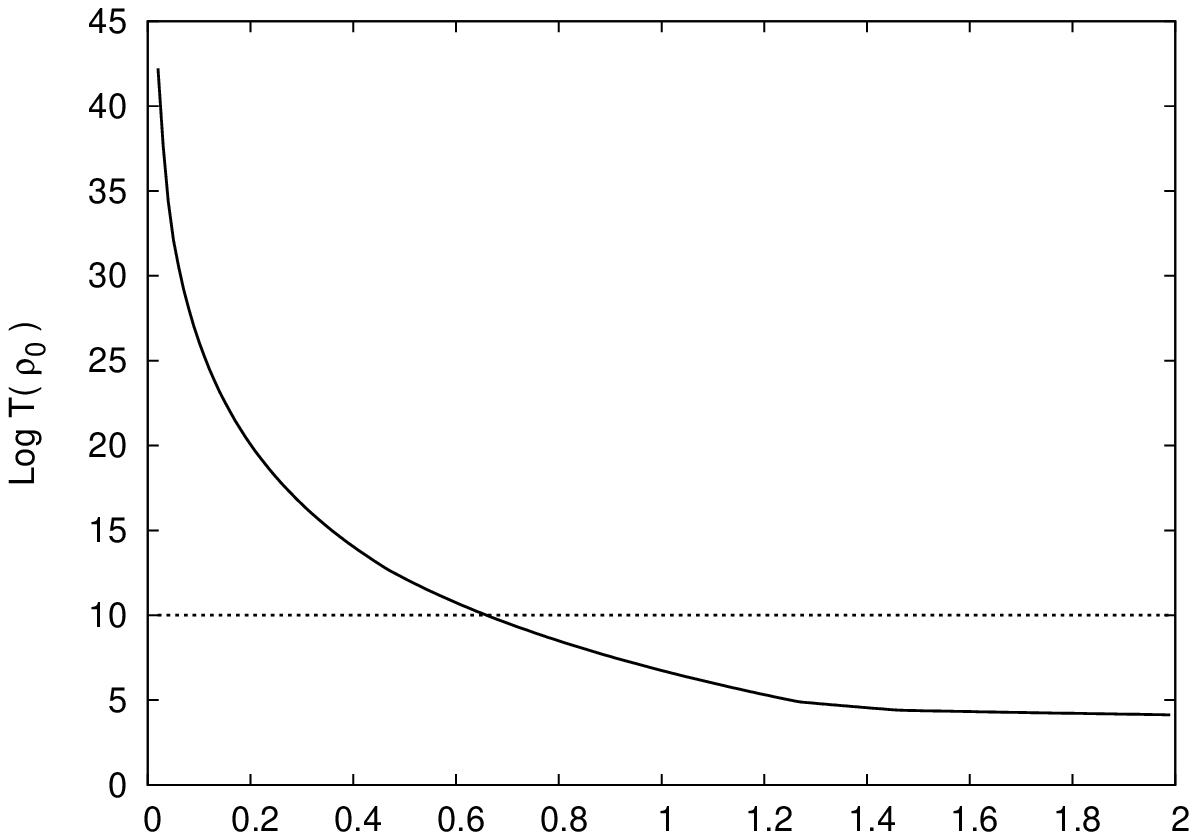,width=14truecm}}
\cr
{\psfig{figure=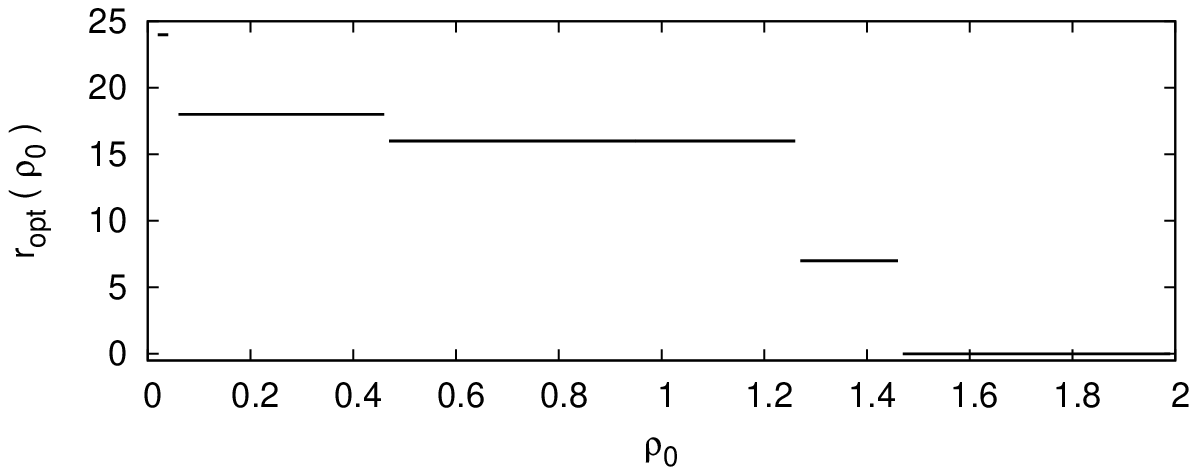,width=14truecm}}
\cr
}}}{The estimated stability time and the optimal truncation order for
the SJSU planar system. The dashed line corresponds to the estimated
age of the Universe.}

\subsection{5.3}{The SJSU system}
As a third application we consider the problem of stability for the
planar secular planetary model including the Sun and  three 
planets, namely Jupiter, Saturn and Uranus.  The aim is evaluate how
long the semi-major axes and the eccentricities of the orbits remain
close to the current value (see~\dbiref{Sansottera-2010}). 

The problem here is much more difficult than in the previous cases.
The Hamiltonian must be expanded in Poincar\'e variables, and is
expressed in action-angle variables for the fast motions and in
Cartesian variables for the slow motions, for a total of 9 polynomial
and 3 trigonometric variables.  The expansion of the
Hamiltonian in these variables clearly is a major task, that has been
handled via computer algebra.

The reduction to the secular problem actually removes the fast
motions, so that we get an equilibrium corresponding to an orbit of
eccentricity zero close to a circular Keplerian one, and a
Hamiltonian expanded in the neighborhood of the equilibrium, which is
still represented as a system of perturbed harmonic oscillators, as in
the cases above.  Thus, after a long preparatory work, we find a
problem similar to the previous one, that can be handled with the same
methods.

The results are represented in fig.~\figref{sjsu}, where we report
again the optimal truncation order and the estimated stability time,
in the same sense as above.  The time unit here is the year, and the
distance is chosen so that $\rho_0=1$ corresponds to the actual
eccentricity of the three planets.  The result is still realistic,
although a stability for a time of the order of the age of Universe
holds only inside a radius corresponding roughly to $70\%$ of the real one.

\par\goodbreak
\blankii
{\parfnt Acknowledgments.}\space
The work of M.~S. is supported by an FSR Incoming Post-doctoral Fellowship of
the Acad\'emie universitaire Louvain, co-funded by the Marie Curie
Actions of the European Commission.

\references

\bye